\documentclass[
 reprint,
superscriptaddress,
nofootinbib,
bibnotes,
 amsmath,amssymb,
 aps,
 prl,
]{revtex4-2}

\usepackage{graphicx}
\usepackage{dcolumn}
\usepackage{bm}
\usepackage[hidelinks]{hyperref}
\usepackage[mathlines]{lineno}
\usepackage[dvipsnames]{xcolor}

\begin{document}

\preprint{APS/123-QED}

\title{Unveiling the physics of the spin-orbit coupling at the surface of a model topological insulator: from theory to experiments}

\author{D.~Puntel} 
 \affiliation{Dipartimento di Fisica, Università degli Studi di Trieste, 34127 Trieste, Italy}

 \author{S.~Peli}
 \affiliation{Elettra - Sincrotrone Trieste S.C.p.A., Strada Statale 14, km 163.5, 34149 Trieste, Italy}
 
 \author{W.~Bronsch}
 \affiliation{Elettra - Sincrotrone Trieste S.C.p.A., Strada Statale 14, km 163.5, 34149 Trieste, Italy}
 
\author{F.~Cilento}%
\affiliation{Elettra - Sincrotrone Trieste S.C.p.A., Strada Statale 14, km 163.5, 34149 Trieste, Italy}%

 \author{H.~Ebert}
 \affiliation{Department Chemie, Ludwig-Maximilians-University M\"unchen, Butenandtstr. 5-11, 81377 München, Germany}
 
 \author{J.~Braun}
 \affiliation{Department Chemie, Ludwig-Maximilians-University M\"unchen, Butenandtstr. 5-11, 81377 München, Germany}
 
 \author{F.~Parmigiani}
 \email{fulvio.parmigiani@elettra.eu}
 \affiliation{Dipartimento di Fisica, Università degli Studi di Trieste, 34127 Trieste, Italy}
 \affiliation{Elettra - Sincrotrone Trieste S.C.p.A., Strada Statale 14, km 163.5, Trieste, Italy}
 \affiliation{International Faculty, University of Cologne, Albertus-Magnus-Platz, 50923 Cologne, Germany}

\begin{abstract}

Spin-orbit interaction affects the band structure of topological insulators beyond the opening of an inverted gap in the bulk bands, and the understanding of its effects on the surface states is of primary importance to access the underlying physics of these exotic states. Here, we propose an \emph{ab initio} approach benchmarked by pump-probe angle-resolved photoelectron spectroscopy data to model the effect of spin-orbit coupling on the surface states of a topological insulator. The critical novelty of our approach lies in the possibility of accounting for a partial transfer of the spin-orbit coupling to the surface states, mediated by the hybridization with the surface resonance states. In topological insulators, the fraction of transferred spin-orbit coupling influences the strength of the hexagonal warping of the surface states, which we use as a telltale of the capability of our model to reproduce the experimental dispersion. The comparison between calculations and measurements, of both the unoccupied and part of the occupied Dirac cone, indicates that the fraction of spin-orbit coupling transferred to the surface states by hybridization with the resonance states is between 70\% and 85\% of its full atomic value. This offers a valuable insight to improve the modeling of surface state properties in topological insulators for both scientific purposes and technological applications. 

\end{abstract}

\maketitle

\section{Introduction}

\begin{figure}[t]
\centering
\includegraphics[width=0.9\columnwidth]{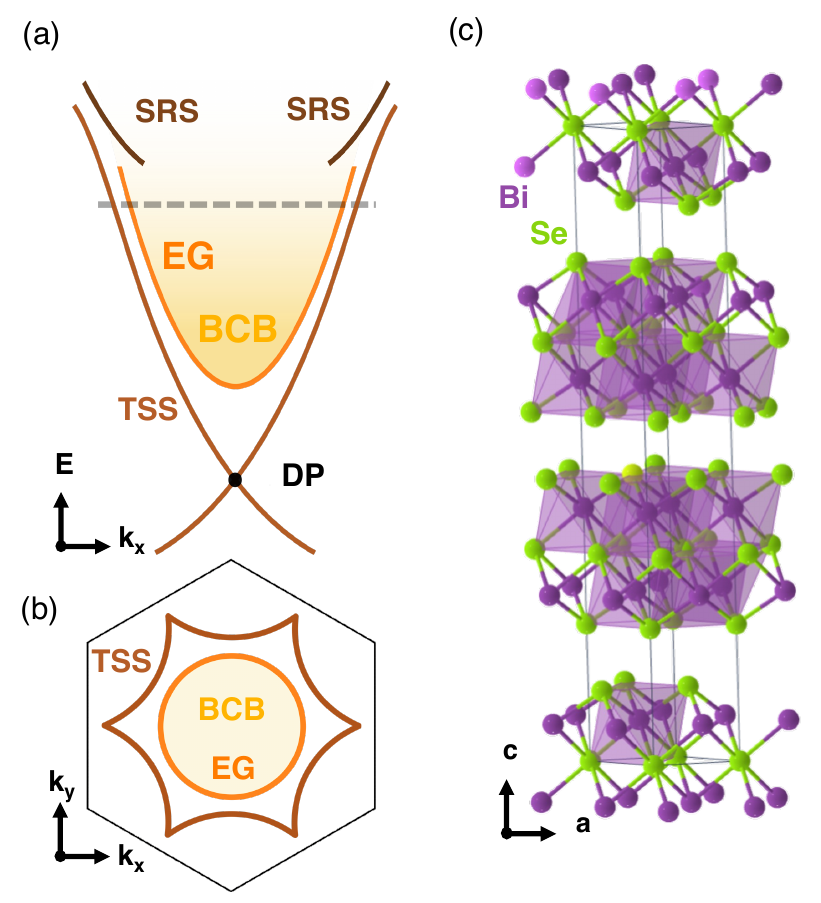}
\caption{(a) Scheme of the dispersion of Bi$_2$Se$_3$. The surface resonance state (SRS, dark brown), the 2D electron gas (EG, orange), the bulk conduction band (BCB, yellow), the topological surface state (TSS, brown) and the Dirac point (DP, black) are indicated. (b) CES cut at the dashed line of panel (b), with the features indicated accordingly. The outer black hexagon indicates the orientation of the first Brillouin zone (not scaled). (c) Conventional unit cell of Bi$_2$Se$_3$ with Bi atoms in purple and Se atoms in green. The quintuple layers are highlighted by purple tetrahedra. This panel was generated using the atomic coordinates from the Materials Project database \cite{Jain2013, MatPrj} through the software VESTA \cite{Momma2008}.
}
\label{Fig_BgTIs}
\end{figure}

Spin-orbit coupling (SOC) in topological insulators (TIs) is responsible for the opening of an inverted bandgap between the valence and the conduction band that endows the band structure with non trivial topology \cite{Kane2005a, Fu2007, Moore2007, Hsieh2008, Hasan2010}. In the surface regime separating the nontrivial insulator and the vacuum, topology arguments dictate the existence of Dirac-like states inside the gap, which are known as topological surface states (TSSs, schematized with brown lines in Fig.~\ref{Fig_BgTIs}(a)) \cite{Zhang2009, Qi2011, Bansil2016}. Since the majority of the technological applications of TIs is based on the electronic properties of the TSS \cite{Roushan2009, Pesin2012, Hamdou2013, Pandey2021, Leppenen2022, Yang2022}, a detailed understanding of the influence of SOC on these states is vital, as it can provide pathways to tailor the properties of the TSS and, therefore, the optical and transport properties of the system. As a recent experimental determination of the e-ph coupling strength in BiSe and BiTe has shown \cite{Huang2023}, a quantitative understanding of the mechanisms that affect the physical properties of TSS in TIs can unlock the gate to control the transport degree of freedom at the surface of these materials. 

In addition to the TSS, the bandstructure of model TIs in the vicinity of the Fermi level comprises several features, summarized in Fig.~\ref{Fig_BgTIs}(a): unoccupied surface resonance states (SRSs, in dark brown) \cite{Cacho2015, Jozwiak2016, Hedayat2021}, a two-dimensional electron gas arising from surface defects (in orange) \cite{Bianchi2010} and the bottom of the conduction band (in shaded yellow), which is typically populated due to unintentional electron doping of the surface \cite{Chen2012}. A deviation of the TSSs dispersion from linearity has been found to lead to the so-called \emph{hexagonal warping} of the constant-energy surfaces (CESs) in the $k_x - k_y$ plane (Fig.~\ref{Fig_BgTIs}(b)) \cite{Kuroda2010, SanchezBarriga2014, Nomura2014, Nurmamat2018}. The hexagonal warping has attracted great attention due to the consequences on the out-of-plane spin polarization of the TSS \cite{Basak2011, Wang2011} and the modification of the scattering properties due to a larger nesting of the Fermi surface \cite{Fu2009}. The phenomenon is most often explained by means of effective models like the $\mathbf{k} \cdot \mathbf{p}$ Hamiltonian \cite{Fu2009, Liu2010}. However, an \emph{ab initio} characterization of the mechanism through which the SOC acts on the surface state dispersion is still lacking. 

On the other hand, this approach was applied to the Rashba surface state of Au(111) in order to investigate the effect of SOC on topologically-trivial surface states from an \emph{ab initio} perspective \cite{Nuber2011}. The comparison with photoemission experiments has shown that the correct value for the Rashba splitting of these states is recovered assuming that only the 50\% of the full SOC of atomic Au is acting on the surface states. 

The evidence of a partial SOC transfer in surface states of simple metals raises the question of whether this mechanism is common also to the surface states of TIs. The two cases are rather different, since topological surface states retain a marked bulk character \emph{via} the hybridization with the SRSs. Hence, at variance with the case of Au(111) Rashba states, the SOC is mediated to the surface states \emph{via} hybridization with the SRSs. 

Our study indicates that it is possible to model the hybridization-mediated transfer of SOC from the bulk to the surface states in the framework of an \emph{ab initio} theory. This coupling allows the transfer to surface states of a fraction of the atomic SOC, similarly to the case of Au(111) Rashba states. Along with other changes in the dispersion, the fraction of transferred SOC influences the strength of the TSS warping, in agreement with the prediction of the $\mathbf{k} \cdot \mathbf{p}$ Hamiltonian. On these bases, we performed one-step photoemission calculations for the prototypical topological insulator Bi$_2$Se$_3$ (Fig.~\ref{Fig_BgTIs}(c)) varying the fraction of SOC transferred to surface states from its full atomic value down to 50\%. To prove the validity of this model, we benchmark our results with pump-probe Angle-Resolved Photoelectron Spectroscopy (ARPES) measurements on Bi$_2$Se$_3$ single crystals. The Dirac point in as-grown Bi$_2$Se$_3$ crystals falls below the Fermi level, for an extent that depends on the amount of substitutional defects in the bulk \cite{Chen2012} and additional aging of the surface \cite{Bianchi2010}. Thanks to this effect, equilibrium ARPES can be capable of detecting evidences of the hexagonal warping \cite{Nomura2014}. However, since the warping increases with the distance from the Dirac point, pump-probe ARPES is advantageous since it allows access to the unoccupied part of the spectrum, where the changes in the TSS dispersion due to warping are more severe \cite{Nurmamat2018}. A match of the calculations with the spectroscopic data, both above and below the Fermi level, is recovered when assuming that the fraction of the SOC transferred to the surface is less than 85\,\%, but in excess of about 70\,\%, hence giving physical consistency to the model we employed to describe the hybridization-mediated partial transfer of SOC to surface states. 

\section{Modeling the spin-orbit coupling mechanism at a solid surface}

In topologically trivial systems, the emergence of surface states is related to the symmetry breaking introduced by the surface itself. The surface properties are best accounted for by a realistic potential barrier, in our case of Rundgren-Malmstr\"om type \cite{Malmstroem1980}, which includes no relativistic effect and is treated as an additional layer on top of the surface. The bulk band structure properties, on the other hand, are determined from the fully-relativistic atomic potentials in the context of the Dirac equation. To determine the coupling between the two, multiple scattering theory must be employed \cite{Braun2018}. This coupling can be allowed to be fractional, meaning that only a fraction of the relativistic properties of the atomic potentials, and thus also of the SOC, is transferred to the surface states. The physical validity of this model is proven by the results obtained on the Rashba splitting of the Au(111) surface state \cite{Nuber2011} and further confirmed by the successful application of the same model to the Shockley state at the Ag/Pt(111) interface \cite{Bendounan2011}. 

Since the existence of the TSSs directly derives from the topologically nontrivial nature of the bulk bandstructure, the bulk character of the TSSs is much more pronounced in TIs than in metal surfaces, with the consequence that the TSS extends for several layers inside the bulk. From the bandstructure point of view, the bulk character is transferred to the TSS by the hybridization with SRS \cite{Cacho2015}. These facts have two important consequences for the physics of SOC at the surface of a topological insulator. First, to determine the effect of the SOC on surface states it is not sufficient to compute the scattering properties between the surface barrier and the atomic potentials, since this approach accounts only for physics localized entirely at the surface, which is not the case for TSS.  Second, the SOC influences the surface states \emph{via} a mechanism that is more complex than the multiple scattering between two potentials, namely the hybridization with the surface resonances, for which no analytical treatment or numerical simulation is presently available. This complexity forces one to model the SOC reduction in an effective way. The pure band structure calculation is performed assuming the full value of the SOC for all layers of the semi-infinite bulk. The atomic potentials are then used as an input for the photoemission calculation, which includes the surface barrier and its coupling to the atomic potentials \emph{via} multiple scattering. In the photoemission step of the calculation, where surface states and resonances are fully taken into account, the SOC of the atoms in the first 20 layers is reduced to a tunable fraction of its atomic value. In this way, the model reproduces a situation where the bulk bandstructure is determined with the full SOC, while only the surface states experience a smaller fraction of SOC, hence effectively modeling the partial SOC transfer through the hybridization with the SRS. 

The modification of the SOC was applied in the spectroscopic calculations in the first 20 layers of the semi-infinite stack of atomic layers, accounting for a reduction of the SOC strength as felt by the surface-related electronic structure. Since Bi atoms are twice as heavy as the Se atoms and thus the SOC is stronger, the SOC reduction is operated on Bi atoms only. The underlying approach focuses on the SOC term of the first-principle Hamiltonian while leaving all other relativistic corrections unchanged \cite{Ebe22}. 

\section{Methods}

The electronic structure calculations were performed self-consistently in the \emph{ab initio} framework of fully relativistic spin-density functional theory, solving the corresponding Dirac equation. The Perdew-Burke-Ernzerhof parametrization was used for the exchange and correlation potentials \cite{Perdew1996}. Furthermore, these calculations were performed for the full SOC. For the spectroscopical analysis, the spin-density matrix formulation of the fully-relativistic one-step model was employed \cite{Braun2014, Braun2018}. The surface of the solid was represented by a Rundgren-Malmstr\"om-type potential \cite{Grass1993}, already successfully applied to the case of Bi$_2$Se$_3$ \cite{Jozwiak2016, Datzer2017}. As this form of the surface barrier implies a $z$-dependent potential, the total photocurrent also includes a surface contribution that explicitly accounts for the dispersion of all the surface features. The photoemission calculations were performed for the photon energy used in the measurements (10.8\,eV, see below) and for the geometry of our experimental setup. To correctly reproduce the relative intensity of the surface spectral features, the matrix elements in the surface region were calculated \cite{Nuber2011}. 

The measurements were performed at the Time-Resolved ARPES endstation of the \mbox{T-ReX} laboratory (Elettra, Trieste). The setup is based on the output of an Yb-doped fiber laser (Coherent Monaco), a fraction of which is used as a pump (1.2\,eV). A cascade of harmonic generation stages in BBO crystals and in Xe gas generates the ninth harmonic of the other part of the beam (10.8\,eV). The incidence angle of the pump and the probe beams on the sample at normal emission is of 30$^{\circ}$, which applies to all the measurements of this work. Further details on the setup are reported elsewhere \cite{Peli2020}. All the measurements were performed at 110\,K and with $s$ probe polarization unless otherwise indicated. 

The pump-probe measurements were performed at the fixed time delay  $\Delta t\sim$1\,ps between the pump and the probe pulses, at which the population in the vicinity of the Fermi level is maximum due to the cascade relaxation mechanism of the charge carriers \cite{Crepaldi2012, Crepaldi2013}. The system was excited with a $p$-polarized pump at a fluence of $\sim$150\,\textmu$\mathrm{J/cm^2}$. 

The CESs were measured with the $p$-polarized probe by rotating the sample in steps of $0.5^{\circ}$ along the tilt axis (perpendicular to the analyzer slit, corresponding to $k_y$ in our axis system for the reciprocal space) and recording an \emph{E~vs.~$k_x$} map for each angle. These maps are stacked together to obtain the full 3D map of the $I(E, k_x, k_y)$ photoemission intensity, which can be visualized by slicing it at constant energy.

\section{Results}
\subsection{\emph{Ab initio} one-step photoemission calculations varying the SOC strength}
\label{Sec_AbInit}

\begin{figure}[t]
\centering
\includegraphics[width=\columnwidth]{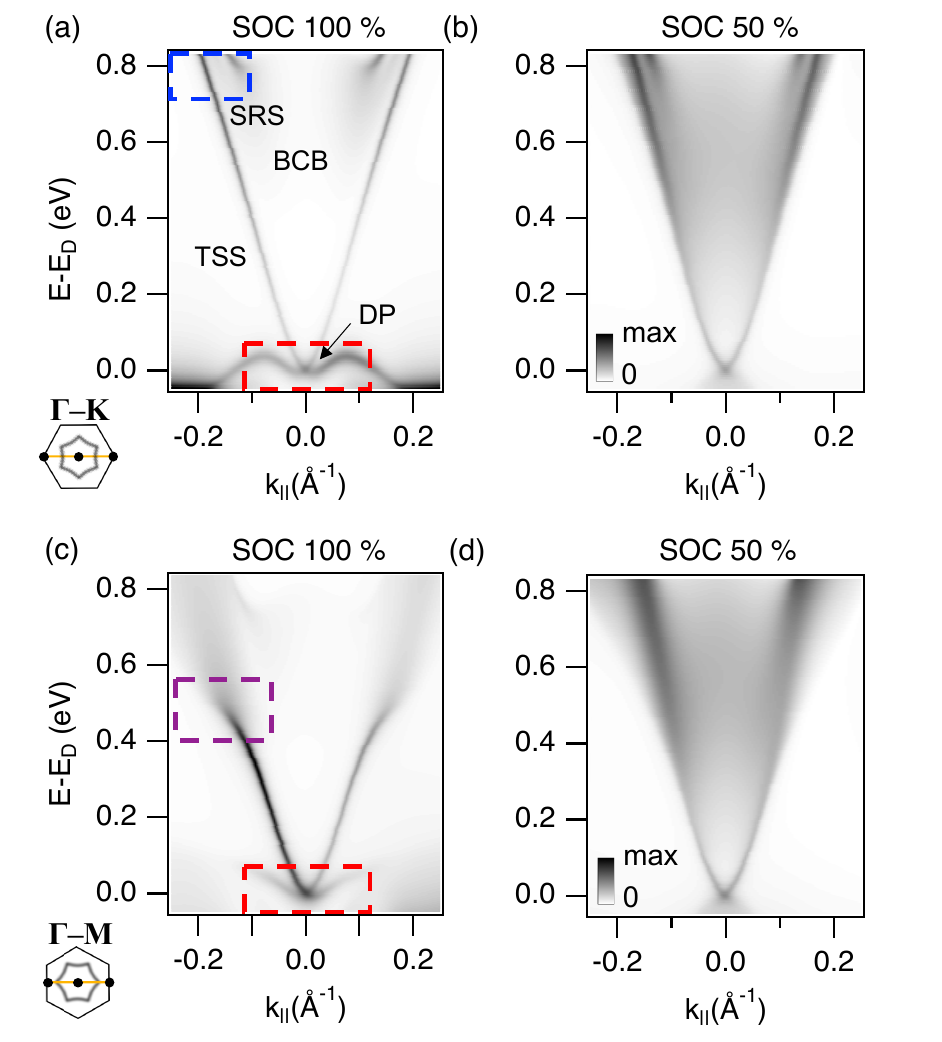}
\caption{
(a),(b) Calculated \emph{E~vs.~k} photoemission intensity distribution along the $\Gamma$K direction for $s$-polarized 10.8\,eV photons, upon varying the SOC strength from 100\% (panel a) to 50\% (panel b). The energy scale is referenced to the energy $E_D$ of the DP. The sketch in the lower left part of panel (a) shows the direction of $k_{\parallel}$ (yellow line) both with respect to the Brillouin zone (outer hexagon) and to the Fermi surface. Panel (a) also indicates the relevant features for our discussion: TSS, topological surface state; BCB, bulk conduction band; SRS, surface resonance state (highlighted by the dashed blue box); DP, Dirac point (dashed red box). (c),(d) Same as (a),(b), but along the $\Gamma$M direction (see low left part of panel (c)). Boxes in panel (c) highlight the point at which the band strongly bends outwards and severely broadens (violet box) and the vicinity of the Dirac point (red box). 
}
\label{Fig_SOCTh}
\end{figure}

To analyze in detail the impact of the SOC on the TSS we scale its strength from 100\,\% of its full atomic value to 50\,\%. Figure~\ref{Fig_SOCTh}(a) shows the calculated photoemission maps along the $\Gamma$K direction for $s$-polarized 10.8\,eV photons, for the cases of 100\% and 50\% SOC (see Supplemental Material \cite{supp}). The energy scale is relative to the Dirac point (DP) and ranges from slightly below the DP to $\sim$800\,meV above it. The linearly dispersing Dirac cone displays the characteristic ``V'' shape above the DP, while close to it the dispersion slightly deviates from exact linearity. The part below the DP instead strongly deviates from the linear dispersion, displaying an ``M''-like shape (dashed red box). No population is observed in between the two branches of the TSS up to $\sim 500$\,meV above the DP. From here to higher energies a small diffuse population is visible, which we attribute to the surface projection of the bulk conduction band (BCB). Around $\sim 800$\,meV above the DP a second sharp feature emerges (dashed blue box). The position and dispersion of this feature is compatible with the SRS mentioned above \cite{Cacho2015}. 

Upon decreasing the SOC strength, several changes occur in the band structure. We first observe that the “M” shape of the dispersion below the DP is smeared out in favor of a monotonic dispersion (panel (d)). The TSS also broadens, with a width that increases with the distance from the DP. Alongside with this, an increase in the BCB population is visible as a gradual change of contrast between the area enclosed by the TSS and that outside. The small intensity of the BCB with respect to the TSS is due to a matrix element effect at this photon energy \cite{Kuroda2010}. This is advantageous since it allows a larger contribution from the TSS, which is the focus of our work. At 50\,\% SOC, however, the BCB intensity is strong and visible down to $\sim$100\,meV above the DP. Apart from the broadening, the dispersion of both the TSS and the SRS along $\Gamma$K undergoes minimal changes, and the TSS remains linear for the whole energy region above the DP. 

The dependence on the SOC of the dispersion along $\Gamma$M is strikingly different. Panel (c) shows the dispersion for the full value of the SOC. Similarly to panel (a), also here the TSS below the DP is bending upwards (dashed red box). The dispersion, while being approximately linear in the energy range between 100 and 300\,meV, is strongly bending outwards around 400\,meV (dashed violet box). We stress that it is this downward bending along $\Gamma$M, along with the preserved linearity along $\Gamma$K, that determines the warping. In the same energy range, the bands start to broaden until they are very diffuse. Also in this case the surface state is visible around $\sim$800\,meV above the DP, although the contrast with respect to the TSS is markedly weaker. Moreover, in panel (d) no intensity appears at $\sim$800\,meV above the DP, where the SRS is expected.

\subsection{Comparing calculations with experiment}

\subsubsection{Dispersion of the occupied band structure}

\begin{figure*}[]
\centering
\includegraphics[width=\textwidth]{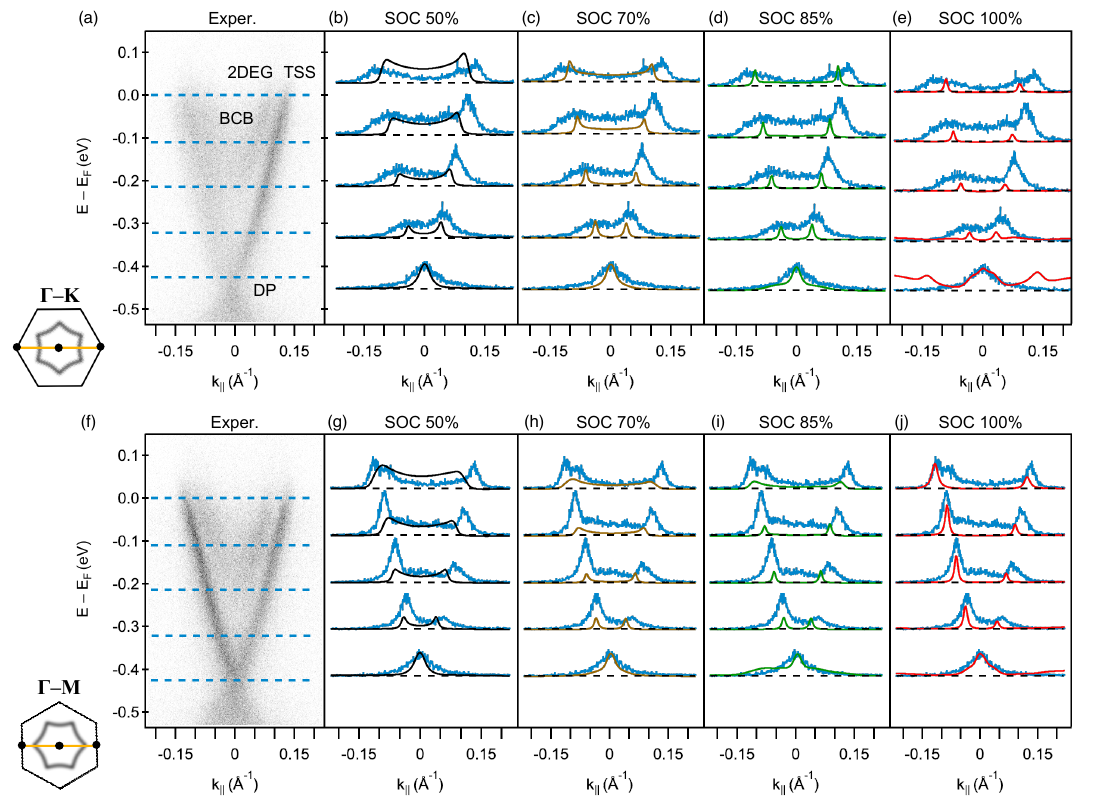}
\caption{(a),(f) Measured band structure along the $\Gamma$K and $\Gamma$M directions, as also shown by the lower parts of panels (a) and (f) . (b-e),(g,j) MDC cuts extracted from the measurement and from the calculations with increasing SOC strength: (b),(g) 50\,\%, (c),(h) 70\,\%, (d),(i) 85\,\%, (e),(j) 100\,\%. The cuts are extracted at the DP, at $E_F$ and at three intermediate energies. Energies are referenced to the experimental $E_F$, \emph{i.e.} shifting down the energy scale of the calculations by $\sim$410\,meV.
}
\label{Fig_UnpGKGM}
\end{figure*}

The occupied band structure is investigated with equilibrium ARPES measurements. Figure~\ref{Fig_UnpGKGM} shows the comparison between the experimental and the theoretical dispersions along the two high-symmetry directions $\Gamma$K (top row) and $\Gamma$M (bottom row). In order to compare the experimental results with the \emph{ab initio} calculations, a common energy scale is needed, since in general the Fermi levels in the calculations and in the experiment might not coincide. The energy scales are thus aligned at the Dirac point, which is an unambiguous feature common to both, and referenced to the experimental Fermi energy $E_F$ (see Supplemental Material \cite{supp}).

In the measured dispersion along $\Gamma$K (Fig.~\ref{Fig_UnpGKGM}(a)), the DP is located $\sim 410$\,meV below $E_F$, hence signaling an $n$-doping of the system due to aging \cite{Hsieh2009a, Bianchi2010}. This is advantageous for our investigation since it allows probing a larger part of the band structure with ARPES. The dispersion below the DP is monotonic and almost linear. Above the DP, the intensity of the two TSS branches becomes asymmetric. While the branch at negative $k$ values (left branch) is weak and broad, the one at positive $k$ values (right branch) is sharply defined. In between the TSS branches, the signature of a 2D electron gas (2DEG), already reported for Bi$_2$Se$_3$ \cite{Bianchi2010}, is visible as a narrow feature at positive $k$. The presence of the 2DEG confirms the aging of the surface also revealed by the relatively large distance between the DP and $E_F$. Also for this feature, the dispersion at negative $k$ is weaker and broader, which makes it hard to distinguish the TSS and the 2DEG. The contribution of the BCB is small due to the unfavorable matrix element cited above \cite{Kuroda2010}. 

The dashed blue lines mark the energies at which the momentum distribution cuts (MDC) have been extracted, \emph{i.e.} at intervals of $\sim 100$\,meV from the DP to $E_F$ and integrated in an energy window of 10\,meV. The MDCs are reported in panels (b-e) in light-blue and compared to the MDCs extracted from the calculated maps at variable SOC strength: (b),(g) 50\,\%, (c),(h) 70\,\%, (d),(i) 85\,\%, (e),(j) 100\,\%. 

Comparison of the dispersion in the vicinity of the DP evidences the inadequacy of the full SOC (red MDC) to account for the measured dispersion, where the TSS shows no ``M''-like shape: its linearity is instead best reproduced at 50\,\% SOC (Fig.~\ref{Fig_SOCTh}(a) and (d)). Above the DP, the MDCs show a discrepancy between experiment and calculations. In particular, the calculated MDCs seem to underestimate the cone aperture, hence predicting a higher band velocity than the one actually observed, and the discrepancy increases with the distance from the DP. This can be explained as a shortcoming of the local spin-density approach, as already reported in \cite{Nuber2011, Aguilera2019, Ponzoni2023}. For the MDCs $\sim 100$\,meV and $\sim 200$\,meV below $E_F$, the 50\,\% SOC seems to give a better agreement, due to the higher population of the BCB with respect to the other SOC values. The intensity measured in this region, however, is more correctly attributable to the 2DEG, since its two branches get close and due to their broadening they partially merge. Overall, the measured relative intensity between the left and right branch of the TSS is well reproduced by all the SOC values, demonstrating that the spectroscopic calculations correctly account for the matrix element effects in our experimental geometry. 

Panel (f) shows the measured dispersion for the $\Gamma$M direction. The intensity distribution is more symmetric than in panel (a), both for the TSS and for the 2DEG. Along this direction also the dispersion in the vicinity of the DP for a SOC of 85\,\% is evidently inadequate, since the bands are bending upwards (see the bump in the MDC at $k\simeq -1\,\mathrm{\AA^{-1}}$). Upon approaching the Fermi level, the difference among the theoretical dispersions becomes more evident. Due to the same overestimation of the band velocity commented above, the 100\,\% SOC seems to agree better with the experimental dispersion at the Fermi level, since the effect of the warping is a detectable reduction of the band velocity along $\Gamma$M. The experimental dispersion, however, is not reflecting the strong warping experienced by the 100\,\% SOC case (Fig.~\ref{Fig_SOCTh}(e)). \\

The comparison between measurements and calculations for the occupied part of the spectrum constitutes a first benchmark for the SOC fraction transferred to the TSS. The 100\,\% SOC calculation does not satisfactorily reproduce the dispersion deduced from the experiments below the DP, for both directions. In addition, the full SOC predicts at $E_F$ a warping larger than the one observed in the experiment. This leads to the first conclusion that the effective fraction of the SOC transferred to the surface states is smaller than the full atomic value. 

\subsubsection{Dispersion of the unoccupied band structure}
\label{SubS_BiSeUnoc}

\begin{figure*}[ht!]
\centering
\includegraphics[width=\textwidth]{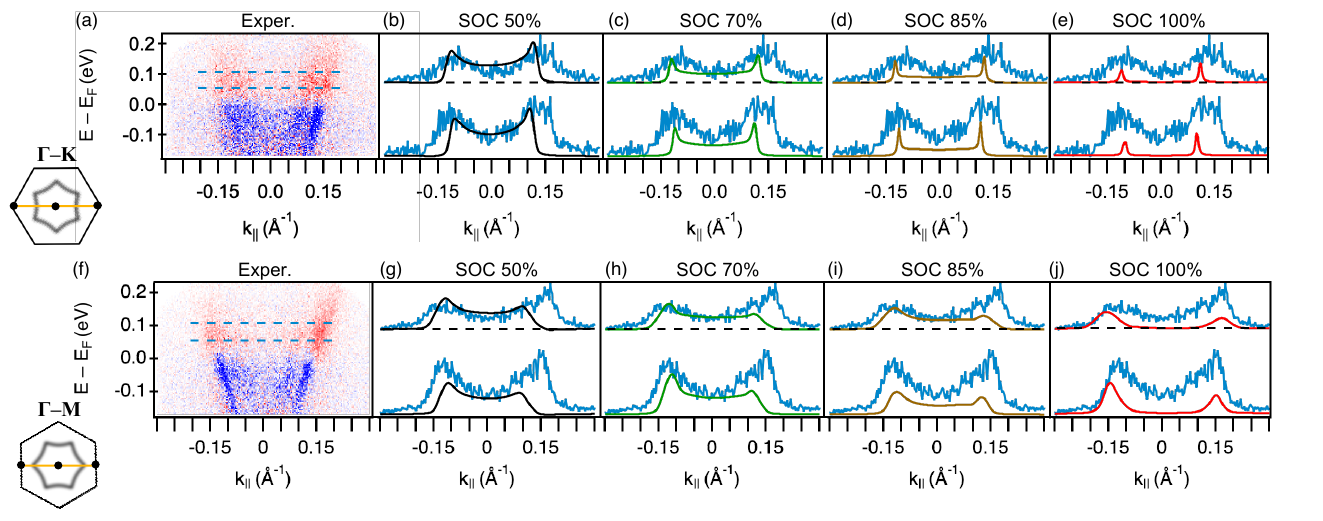}
\caption{Pumped counterpart of Fig.~\ref{Fig_UnpGKGM}. Top (bottom) row refers to the $\Gamma$K ($\Gamma$M) direction. The experimental dispersion is reported in (a) and (g) as a difference map between the one acquired at $\Delta t = 1$\,ps and the equilibrium one, to evidence the population of the band structure above $E_F$. (b-e),(g,j) MDC cuts extracted from the measurement and from the calculations with increasing SOC strength: (b),(g) 50\,\%, (c),(h) 70\,\%, (d),(i) 85\,\%, (e),(j) 100\,\%. The cuts are extracted 55\,meV and 110\,meV above $E_F$. Energies are referenced to the experimental $E_F$, \emph{i.e.} shifting down the energy scale of the calculations by $\sim$410\,meV.
}
\label{Fig_PmpGKGM}
\end{figure*}

Since the warping increases with the distance from the DP \cite{Nurmamat2018}, access to larger portions of the band structure above $E_F$ is beneficial to compare the experiments and the calculations in an energy interval where the influence of the SOC on the warping is more pronounced. Figure~\ref{Fig_PmpGKGM} shows the comparison between measured and calculated photoemission maps following the same arrangement of Figure~\ref{Fig_UnpGKGM}. The measured maps are reported as a difference between the pumped map and the equilibrium one, as customary in time-resolved ARPES measurements to emphasize the change in the photoemission intensity. The measurement along $\Gamma$K (panel (a)) shows that the linear dispersion of the TSS is also observed above $E_F$. The branch at negative $k$ values remains broad and weaker, but the one at positive $k$ values is well distinct from the 2DEG and allows discerning that the latter disperses roughly parallel to the TSS. Accessing the dispersion above $E_F$ is also beneficial for circumventing the unfavorable coincidence of a broad 2DEG and a small BCB contribution to the intensity. As discussed above for the region close to $E_F$, the parabolic dispersion of the 2DEG causes its branches to be well separated in momentum, hence allowing the attribution of the intensity detected in between to the sole BCB. This is better visible in the MDC cuts taken at $\sim$55\,meV and $\sim$110\,meV above $E_F$ and shown in subsequent panels. The calculation for 50\,\% SOC (b) clearly shows an excessive intensity in correspondence of the BCB. The 70\,\% and 85\,\% SOC cases, (c) and (d), are instead closer to the data for the experimental dispersion, while the calculation for 100\,\% SOC in (e) shows no significant BCB intensity. 

The measurement along $\Gamma$M is shown in panel (f). The right branches of the TSS and of the 2DEG display a larger intensity, which is also reflected in the presence of a red signal up to higher energies. This might be due to a delay-dependent matrix element effect \cite{Boschini2020}, although the depopulation of the TSS below $E_F$ is more symmetric. The right branch clearly shows that the dispersion is deviating from linearity. Incidentally, the 2DEG population drastically decreases beyond $\sim$100\,meV above $E_F$, at variance with what is observed along $\Gamma$K. The intensity of the BCB population instead follows a behavior similar to the other direction, and is well reproduced for 70\,\% and 85\,\% SOC, as proven by the MDCs in panel (h). The magnitude of the warping agrees well with the prediction for the 85\,\% and even 100\,\% of the SOC. Since for lower SOC values the TSS is severely smearing out at higher energies, also the sharpness of the features is better accounted for at higher SOC values. \\

The inspection of the pumped \emph{E~vs.~k} maps proves that the intensity from the BCB is small compared to the TSS and the 2DEG. It also allowed to confirm that the dispersion along $\Gamma$K is linear also above $E_F$. The dispersion along $\Gamma$M instead shows a strong warping that is compatible with the highest values of SOC.

\subsubsection{Comparison of constant-energy surfaces}

\begin{figure*}[ht!]
\centering
\includegraphics[width=\textwidth]{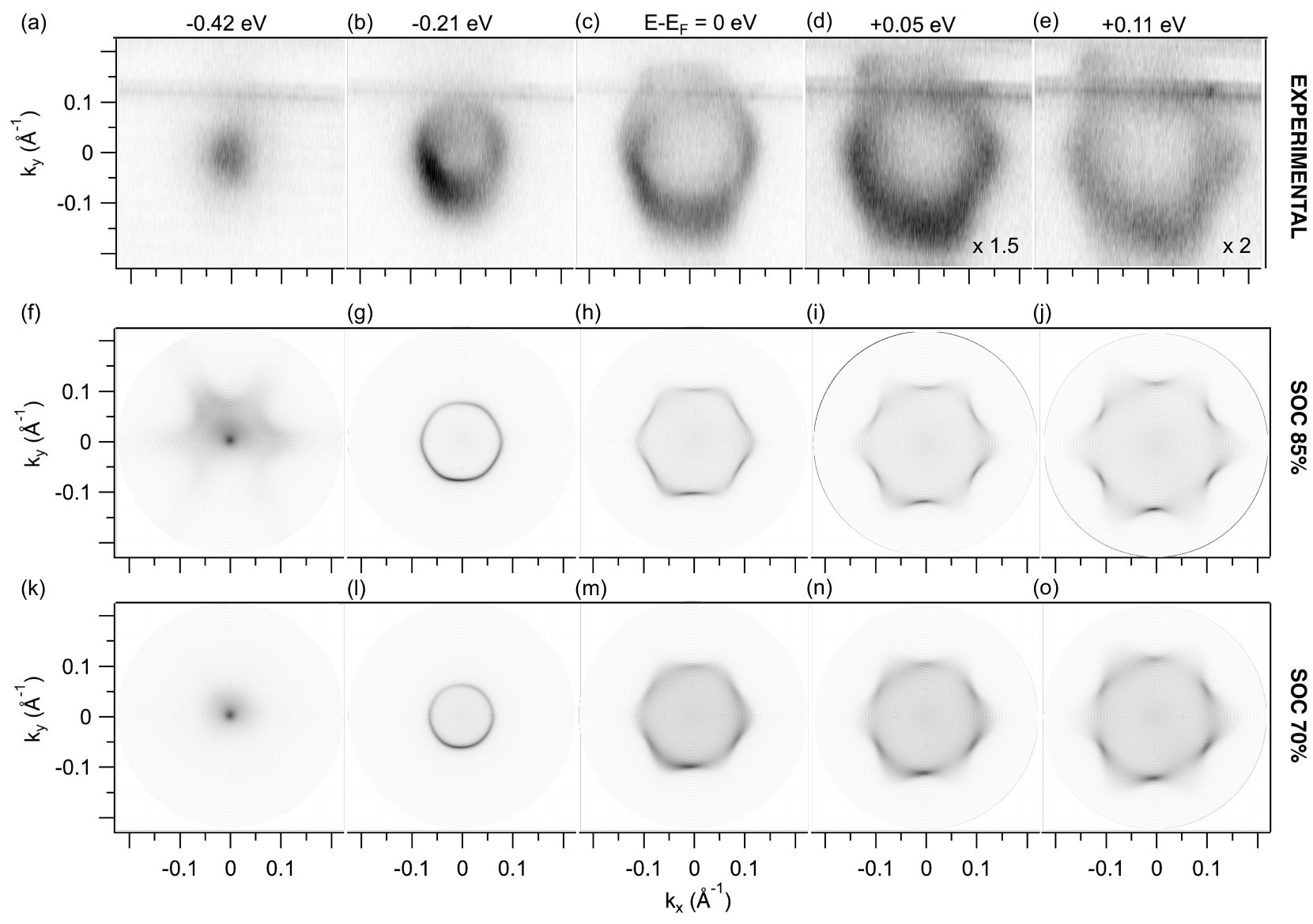}
\caption{
(a-e) Experimental CES for the selected values of energy indicated above: (a) at the DP, (b) between the DP and $E_F$, (c) at $E_F$, (d) 0.05\,meV and (e) 0.11\,meV above $E_F$ ($\Delta t =1$\,ps). Energies are referenced to the experimental $E_F$, \emph{i.e.} shifting down the energy scale of the calculations by $\sim$410\,meV. The intensity of (d) and (e) the last two is multiplied by a factor of 1.5 and 2 for better visibility. 
(f-j) CESs calculated for 85\,\% SOC at the same energies. 
(k-o) CESs calculated for 70\,\% SOC at the same energies. 
}
\label{Fig_UnpCES}
\end{figure*}

The inspection of CESs makes it easier, with respect to the \emph{E~vs.~k} dispersion, to visualize the symmetry and the degree of deformation of the CESs, so to evaluate which strength of the SOC is more appropriate in describing the shape of the experimental CESs. 

The result of this procedure is shown in Fig.~\ref{Fig_UnpCES}(a-e) for the CES measured with $p$-polarized light. The cuts are taken at five energies: at the DP, between the DP and $E_F$, at $E_F$, and at the two energies above $E_F$ already considered in Fig.~\ref{Fig_PmpGKGM}. The TSS contour at the DP (panel (a)) is circular as expected from the vertex of a cone broadened by the experimental resolution. No intensity is detected outside this circular area, apart from a sharp line at $k_y \sim 0.12\,\AA^{-1}$. This spurious intensity is probably the result of the probe beam impinging on a flake edge during the tilt angle scan. The cut at -210\,meV (panel (b)) has a round shape with some larger broadening at negative $k_y$ values, where also the photoemission intensity is larger. The TSS contour at $E_F$ (panel (c)) is also broader at negative $k_y$. At this energy there is a clear deformation towards a hexagon. The small discontinuity at positive $k_y$ values is likely due to a change in the position of the probe on the sample due to the angular motion of the latter, as suggested by the strong spurious strike of intensity commented above. The CES at 50\,meV above $E_F$ (panel (d)) shows a larger deformation of the TSS, going towards the snowflake-like shape. The case of panel (e) is analogous, with the deformation particularly visible on the right edge of the snowflake at $k_x \sim 0$ and in the top left edge. We also emphasize that the contrast between the TSS and the region inside it confirms that the BCB photoemission intensity is very weak. 

The CESs were calculated for the sole SOC values of 70\,\% and 85\,\%, \emph{i.e.} those that yield the best agreement with the experimental \emph{E~vs.~k} data. Panels (f-j) show the 85\,\% SOC case. The ``M'' shaped dispersion commented in Fig.~\ref{Fig_UnpGKGM}(f),(i) close to the DP here appears as a star-like contour of intensity, in contrast with the experimental observation in (a). Panel (g), showing an already deformed contour, agrees well with the experimental counterpart of panel (b), notwithstanding the broadening of the latter. The following CESs (h-j) also show a good agreement with the experiment, in particular correctly accounting for the intensity asymmetries. In fact, the most intense regions of the hexagon are towards negative $k_y$ values. 

A better agreement is found for 70\,\% SOC (k-o). The DP (k) is properly reproduced as a small circular area. The CES at -210\,meV shown in (l) is still circular, in partial disagreement with the measurement in (b), and becomes hexagonal only at $E_F$ (m). Another shortcoming of this SOC value is that the warping is less severe, and thus the size of the contours above $E_F$ (n-o) is slightly smaller than what actually measured. Also at this SOC the intensity asymmetry is  reproduced, being the CES more intense in the negative $k_y$ portion. We stress the fact that for both SOC values, the snowflake deformation above $E_F$ comes along with an intense smearing of the TSS, as already commented for Fig.~\ref{Fig_PmpGKGM}(f), with the result that the tips of the snowflake along $\Gamma$M are almost vanishing. 
\\

The comparison between the measured and the calculated CESs gives indications in agreement with what is found from the analysis of the \emph{E~vs.~k} maps. The shape of the CES close to the DP suggests a SOC value lower that 85\,\%. On the other hand, the hexagonal deformation of the 70\,\% SOC occurs closer to $E_F$ than the experiment suggests, and the size of the CES contour above $E_F$ is smaller, pointing to a value of the SOC closer to 85\,\%. The CESs also further show that the calculations correctly account for matrix element effects. 

\section{Conclusions}

The present study combines state-of-art \emph{ab initio} one-step photoemission calculations with equilibrium and pump-probe ARPES measurements, to investigate the physics of the SOC at the surface of the model TI Bi$_2$Se$_3$. Our study proves that the influence of the SOC on TSSs can be modeled within a theoretical framework capable of taking into account the partial transfer of the atomic SOC to the surface states, which occurs \emph{via} hybridization with the surface resonance states. 

The calculations show that the most severe changes in the TSS dispersion upon changing the fraction of transferred SOC occur along the $\Gamma$M direction, where the TSS branches bend outwards for higher SOC values. At the same time, they remain linear at all SOC strengths along $\Gamma$K. Therefore, a stronger SOC induces a stronger warping, in qualitative agreement with the prediction of the $\mathbf{k} \cdot \mathbf{p}$ model. 

The comparison of the SOC-dependent calculations for the surface states with the ARPES measurements shows that the highest values of SOC are incompatible with the experimental evidence in the vicinity of the Dirac point. In addition, the bare atomic SOC predicts a warping in excess of the one found in the experiment. Instead, the unoccupied part of the spectrum shows that the warping is compatible with the highest values of SOC, hence indicating a transferred SOC fraction larger than 70\%. The comparison between measured and calculated CESs confirms that a SOC value between 70\% and 85\% correctly accounts for the shape and intensity distribution of the CESs. 

In conclusion, the agreement of our ARPES data with the calculations performed for a reduced value of the SOC proves that our model correctly accounts for the physics of the transfer of SOC to the surface states of a topological insulator. Moreover, it allows estimating the fraction of transferred SOC between 70\% and 85\% of its bare atomic value. This indicates that, in order to correctly model the physics of topological surface states, the hybridization with surface resonance states is a crucial ingredient. This accounts for the proven weak coupling between bulk and surface states and offers a way to model the partial SOC transfer from the former to the latter, hence leading towards a clearer understanding of the electronic mechanisms that ultimately shapes the surface band structure of Bi$_2$Se$_3$.


\begin{thebibliography}{48}%
\makeatletter
\providecommand \@ifxundefined [1]{%
 \@ifx{#1\undefined}
}%
\providecommand \@ifnum [1]{%
 \ifnum #1\expandafter \@firstoftwo
 \else \expandafter \@secondoftwo
 \fi
}%
\providecommand \@ifx [1]{%
 \ifx #1\expandafter \@firstoftwo
 \else \expandafter \@secondoftwo
 \fi
}%
\providecommand \natexlab [1]{#1}%
\providecommand \enquote  [1]{``#1''}%
\providecommand \bibnamefont  [1]{#1}%
\providecommand \bibfnamefont [1]{#1}%
\providecommand \citenamefont [1]{#1}%
\providecommand \href@noop [0]{\@secondoftwo}%
\providecommand \href [0]{\begingroup \@sanitize@url \@href}%
\providecommand \@href[1]{\@@startlink{#1}\@@href}%
\providecommand \@@href[1]{\endgroup#1\@@endlink}%
\providecommand \@sanitize@url [0]{\catcode `\\12\catcode `\$12\catcode
  `\&12\catcode `\#12\catcode `\^12\catcode `\_12\catcode `\%12\relax}%
\providecommand \@@startlink[1]{}%
\providecommand \@@endlink[0]{}%
\providecommand \url  [0]{\begingroup\@sanitize@url \@url }%
\providecommand \@url [1]{\endgroup\@href {#1}{\urlprefix }}%
\providecommand \urlprefix  [0]{URL }%
\providecommand \Eprint [0]{\href }%
\providecommand \doibase [0]{https://doi.org/}%
\providecommand \selectlanguage [0]{\@gobble}%
\providecommand \bibinfo  [0]{\@secondoftwo}%
\providecommand \bibfield  [0]{\@secondoftwo}%
\providecommand \translation [1]{[#1]}%
\providecommand \BibitemOpen [0]{}%
\providecommand \bibitemStop [0]{}%
\providecommand \bibitemNoStop [0]{.\EOS\space}%
\providecommand \EOS [0]{\spacefactor3000\relax}%
\providecommand \BibitemShut  [1]{\csname bibitem#1\endcsname}%
\let\auto@bib@innerbib\@empty
\bibitem [{\citenamefont {Kane}\ and\ \citenamefont {Mele}(2005)}]{Kane2005a}%
  \BibitemOpen
  \bibfield  {author} {\bibinfo {author} {\bibfnamefont {C.~L.}\ \bibnamefont
  {Kane}}\ and\ \bibinfo {author} {\bibfnamefont {E.~J.}\ \bibnamefont
  {Mele}},\ }\bibfield  {title} {\bibinfo {title} {Quantum spin {Hall} effect
  in graphene},\ }\href {https://doi.org/10.1103/physrevlett.95.226801}
  {\bibfield  {journal} {\bibinfo  {journal} {Physical Review Letters}\
  }\textbf {\bibinfo {volume} {95}},\ \bibinfo {pages} {226801} (\bibinfo
  {year} {2005})}\BibitemShut {NoStop}%
\bibitem [{\citenamefont {Fu}\ \emph {et~al.}(2007)\citenamefont {Fu},
  \citenamefont {Kane},\ and\ \citenamefont {Mele}}]{Fu2007}%
  \BibitemOpen
  \bibfield  {author} {\bibinfo {author} {\bibfnamefont {L.}~\bibnamefont
  {Fu}}, \bibinfo {author} {\bibfnamefont {C.~L.}\ \bibnamefont {Kane}},\ and\
  \bibinfo {author} {\bibfnamefont {E.~J.}\ \bibnamefont {Mele}},\ }\bibfield
  {title} {\bibinfo {title} {Topological insulators in three dimensions},\
  }\href {https://doi.org/10.1103/physrevlett.98.106803} {\bibfield  {journal}
  {\bibinfo  {journal} {Physical Review Letters}\ }\textbf {\bibinfo {volume}
  {98}},\ \bibinfo {pages} {106803} (\bibinfo {year} {2007})}\BibitemShut
  {NoStop}%
\bibitem [{\citenamefont {Moore}\ and\ \citenamefont
  {Balents}(2007)}]{Moore2007}%
  \BibitemOpen
  \bibfield  {author} {\bibinfo {author} {\bibfnamefont {J.~E.}\ \bibnamefont
  {Moore}}\ and\ \bibinfo {author} {\bibfnamefont {L.}~\bibnamefont
  {Balents}},\ }\bibfield  {title} {\bibinfo {title} {Topological invariants of
  time-reversal-invariant band structures},\ }\href
  {https://doi.org/10.1103/physrevb.75.121306} {\bibfield  {journal} {\bibinfo
  {journal} {Physical Review B}\ }\textbf {\bibinfo {volume} {75}},\ \bibinfo
  {pages} {121306} (\bibinfo {year} {2007})}\BibitemShut {NoStop}%
\bibitem [{\citenamefont {Hsieh}\ \emph {et~al.}(2008)\citenamefont {Hsieh},
  \citenamefont {Qian}, \citenamefont {Wray}, \citenamefont {Xia},
  \citenamefont {Hor}, \citenamefont {Cava},\ and\ \citenamefont
  {Hasan}}]{Hsieh2008}%
  \BibitemOpen
  \bibfield  {author} {\bibinfo {author} {\bibfnamefont {D.}~\bibnamefont
  {Hsieh}}, \bibinfo {author} {\bibfnamefont {D.}~\bibnamefont {Qian}},
  \bibinfo {author} {\bibfnamefont {L.}~\bibnamefont {Wray}}, \bibinfo {author}
  {\bibfnamefont {Y.}~\bibnamefont {Xia}}, \bibinfo {author} {\bibfnamefont
  {Y.~S.}\ \bibnamefont {Hor}}, \bibinfo {author} {\bibfnamefont {R.~J.}\
  \bibnamefont {Cava}},\ and\ \bibinfo {author} {\bibfnamefont {M.~Z.}\
  \bibnamefont {Hasan}},\ }\bibfield  {title} {\bibinfo {title} {A topological
  {Dirac} insulator in a quantum spin {Hall} phase},\ }\href
  {https://doi.org/10.1038/nature06843} {\bibfield  {journal} {\bibinfo
  {journal} {Nature}\ }\textbf {\bibinfo {volume} {452}},\ \bibinfo {pages}
  {970} (\bibinfo {year} {2008})}\BibitemShut {NoStop}%
\bibitem [{\citenamefont {Hasan}\ and\ \citenamefont {Kane}(2010)}]{Hasan2010}%
  \BibitemOpen
  \bibfield  {author} {\bibinfo {author} {\bibfnamefont {M.~Z.}\ \bibnamefont
  {Hasan}}\ and\ \bibinfo {author} {\bibfnamefont {C.~L.}\ \bibnamefont
  {Kane}},\ }\bibfield  {title} {\bibinfo {title} {\emph{Colloquium}:
  Topological insulators},\ }\href {https://doi.org/10.1103/revmodphys.82.3045}
  {\bibfield  {journal} {\bibinfo  {journal} {Reviews of Modern Physics}\
  }\textbf {\bibinfo {volume} {82}},\ \bibinfo {pages} {3045} (\bibinfo {year}
  {2010})}\BibitemShut {NoStop}%
\bibitem [{\citenamefont {Zhang}\ \emph {et~al.}(2009)\citenamefont {Zhang},
  \citenamefont {Liu}, \citenamefont {Qi}, \citenamefont {Dai}, \citenamefont
  {Fang},\ and\ \citenamefont {Zhang}}]{Zhang2009}%
  \BibitemOpen
  \bibfield  {author} {\bibinfo {author} {\bibfnamefont {H.}~\bibnamefont
  {Zhang}}, \bibinfo {author} {\bibfnamefont {C.-X.}\ \bibnamefont {Liu}},
  \bibinfo {author} {\bibfnamefont {X.-L.}\ \bibnamefont {Qi}}, \bibinfo
  {author} {\bibfnamefont {X.}~\bibnamefont {Dai}}, \bibinfo {author}
  {\bibfnamefont {Z.}~\bibnamefont {Fang}},\ and\ \bibinfo {author}
  {\bibfnamefont {S.-C.}\ \bibnamefont {Zhang}},\ }\bibfield  {title} {\bibinfo
  {title} {Topological insulators in {Bi2Se3, Bi2Te3 and Sb2Te3} with a single
  {Dirac} cone on the surface},\ }\href {https://doi.org/10.1038/nphys1270}
  {\bibfield  {journal} {\bibinfo  {journal} {Nature Physics}\ }\textbf
  {\bibinfo {volume} {5}},\ \bibinfo {pages} {438} (\bibinfo {year}
  {2009})}\BibitemShut {NoStop}%
\bibitem [{\citenamefont {Qi}\ and\ \citenamefont {Zhang}(2011)}]{Qi2011}%
  \BibitemOpen
  \bibfield  {author} {\bibinfo {author} {\bibfnamefont {X.-L.}\ \bibnamefont
  {Qi}}\ and\ \bibinfo {author} {\bibfnamefont {S.-C.}\ \bibnamefont {Zhang}},\
  }\bibfield  {title} {\bibinfo {title} {Topological insulators and
  superconductors},\ }\href {https://doi.org/10.1103/revmodphys.83.1057}
  {\bibfield  {journal} {\bibinfo  {journal} {Reviews of Modern Physics}\
  }\textbf {\bibinfo {volume} {83}},\ \bibinfo {pages} {1057} (\bibinfo {year}
  {2011})}\BibitemShut {NoStop}%
\bibitem [{\citenamefont {Bansil}\ \emph {et~al.}(2016)\citenamefont {Bansil},
  \citenamefont {Lin},\ and\ \citenamefont {Das}}]{Bansil2016}%
  \BibitemOpen
  \bibfield  {author} {\bibinfo {author} {\bibfnamefont {A.}~\bibnamefont
  {Bansil}}, \bibinfo {author} {\bibfnamefont {H.}~\bibnamefont {Lin}},\ and\
  \bibinfo {author} {\bibfnamefont {T.}~\bibnamefont {Das}},\ }\bibfield
  {title} {\bibinfo {title} {\emph{Colloquium}: Topological band theory},\
  }\href {https://doi.org/10.1103/revmodphys.88.021004} {\bibfield  {journal}
  {\bibinfo  {journal} {Reviews of Modern Physics}\ }\textbf {\bibinfo {volume}
  {88}},\ \bibinfo {pages} {021004} (\bibinfo {year} {2016})}\BibitemShut
  {NoStop}%
\bibitem [{\citenamefont {Roushan}\ \emph {et~al.}(2009)\citenamefont
  {Roushan}, \citenamefont {Seo}, \citenamefont {Parker}, \citenamefont {Hor},
  \citenamefont {Hsieh}, \citenamefont {Qian}, \citenamefont {Richardella},
  \citenamefont {Hasan}, \citenamefont {Cava},\ and\ \citenamefont
  {Yazdani}}]{Roushan2009}%
  \BibitemOpen
  \bibfield  {author} {\bibinfo {author} {\bibfnamefont {P.}~\bibnamefont
  {Roushan}}, \bibinfo {author} {\bibfnamefont {J.}~\bibnamefont {Seo}},
  \bibinfo {author} {\bibfnamefont {C.~V.}\ \bibnamefont {Parker}}, \bibinfo
  {author} {\bibfnamefont {Y.~S.}\ \bibnamefont {Hor}}, \bibinfo {author}
  {\bibfnamefont {D.}~\bibnamefont {Hsieh}}, \bibinfo {author} {\bibfnamefont
  {D.}~\bibnamefont {Qian}}, \bibinfo {author} {\bibfnamefont {A.}~\bibnamefont
  {Richardella}}, \bibinfo {author} {\bibfnamefont {M.~Z.}\ \bibnamefont
  {Hasan}}, \bibinfo {author} {\bibfnamefont {R.~J.}\ \bibnamefont {Cava}},\
  and\ \bibinfo {author} {\bibfnamefont {A.}~\bibnamefont {Yazdani}},\
  }\bibfield  {title} {\bibinfo {title} {Topological surface states protected
  from backscattering by chiral spin texture},\ }\href
  {https://doi.org/10.1038/nature08308} {\bibfield  {journal} {\bibinfo
  {journal} {Nature}\ }\textbf {\bibinfo {volume} {460}},\ \bibinfo {pages}
  {1106} (\bibinfo {year} {2009})}\BibitemShut {NoStop}%
\bibitem [{\citenamefont {Pesin}\ and\ \citenamefont
  {MacDonald}(2012)}]{Pesin2012}%
  \BibitemOpen
  \bibfield  {author} {\bibinfo {author} {\bibfnamefont {D.}~\bibnamefont
  {Pesin}}\ and\ \bibinfo {author} {\bibfnamefont {A.~H.}\ \bibnamefont
  {MacDonald}},\ }\bibfield  {title} {\bibinfo {title} {Spintronics and
  pseudospintronics in graphene and topological insulators},\ }\href
  {https://doi.org/10.1038/nmat3305} {\bibfield  {journal} {\bibinfo  {journal}
  {Nature Materials}\ }\textbf {\bibinfo {volume} {11}},\ \bibinfo {pages}
  {409} (\bibinfo {year} {2012})}\BibitemShut {NoStop}%
\bibitem [{\citenamefont {Hamdou}\ \emph {et~al.}(2013)\citenamefont {Hamdou},
  \citenamefont {Gooth}, \citenamefont {Dorn}, \citenamefont {Pippel},\ and\
  \citenamefont {Nielsch}}]{Hamdou2013}%
  \BibitemOpen
  \bibfield  {author} {\bibinfo {author} {\bibfnamefont {B.}~\bibnamefont
  {Hamdou}}, \bibinfo {author} {\bibfnamefont {J.}~\bibnamefont {Gooth}},
  \bibinfo {author} {\bibfnamefont {A.}~\bibnamefont {Dorn}}, \bibinfo {author}
  {\bibfnamefont {E.}~\bibnamefont {Pippel}},\ and\ \bibinfo {author}
  {\bibfnamefont {K.}~\bibnamefont {Nielsch}},\ }\bibfield  {title} {\bibinfo
  {title} {{Surface state dominated transport in topological insulator
  Bi$_2$Te$_3$ nanowires}},\ }\href {https://doi.org/10.1063/1.4829748}
  {\bibfield  {journal} {\bibinfo  {journal} {Applied Physics Letters}\
  }\textbf {\bibinfo {volume} {103}},\ \bibinfo {pages} {193107} (\bibinfo
  {year} {2013})}\BibitemShut {NoStop}%
\bibitem [{\citenamefont {Pandey}\ \emph {et~al.}(2021)\citenamefont {Pandey},
  \citenamefont {Yadav}, \citenamefont {Kaur}, \citenamefont {Singh},
  \citenamefont {Gupta},\ and\ \citenamefont {Husale}}]{Pandey2021}%
  \BibitemOpen
  \bibfield  {author} {\bibinfo {author} {\bibfnamefont {A.}~\bibnamefont
  {Pandey}}, \bibinfo {author} {\bibfnamefont {R.}~\bibnamefont {Yadav}},
  \bibinfo {author} {\bibfnamefont {M.}~\bibnamefont {Kaur}}, \bibinfo {author}
  {\bibfnamefont {P.}~\bibnamefont {Singh}}, \bibinfo {author} {\bibfnamefont
  {A.}~\bibnamefont {Gupta}},\ and\ \bibinfo {author} {\bibfnamefont
  {S.}~\bibnamefont {Husale}},\ }\bibfield  {title} {\bibinfo {title} {High
  performing flexible optoelectronic devices using thin films of topological
  insulator},\ }\bibfield  {journal} {\bibinfo  {journal} {Scientific Reports}\
  }\textbf {\bibinfo {volume} {11}},\ \href
  {https://doi.org/10.1038/s41598-020-80738-8} {10.1038/s41598-020-80738-8}
  (\bibinfo {year} {2021})\BibitemShut {NoStop}%
\bibitem [{\citenamefont {Leppenen}\ and\ \citenamefont
  {Golub}(2022)}]{Leppenen2022}%
  \BibitemOpen
  \bibfield  {author} {\bibinfo {author} {\bibfnamefont {N.~V.}\ \bibnamefont
  {Leppenen}}\ and\ \bibinfo {author} {\bibfnamefont {L.~E.}\ \bibnamefont
  {Golub}},\ }\bibfield  {title} {\bibinfo {title} {Nonlinear optical
  absorption and photocurrents in topological insulators},\ }\href
  {https://doi.org/10.1103/PhysRevB.105.115306} {\bibfield  {journal} {\bibinfo
   {journal} {Phys. Rev. B}\ }\textbf {\bibinfo {volume} {105}},\ \bibinfo
  {pages} {115306} (\bibinfo {year} {2022})}\BibitemShut {NoStop}%
\bibitem [{\citenamefont {Yang}\ \emph {et~al.}(2022)\citenamefont {Yang},
  \citenamefont {Zhou},\ and\ \citenamefont {Wang}}]{Yang2022}%
  \BibitemOpen
  \bibfield  {author} {\bibinfo {author} {\bibfnamefont {M.}~\bibnamefont
  {Yang}}, \bibinfo {author} {\bibfnamefont {H.}~\bibnamefont {Zhou}},\ and\
  \bibinfo {author} {\bibfnamefont {J.}~\bibnamefont {Wang}},\ }\bibfield
  {title} {\bibinfo {title} {Topological insulators photodetectors:
  Preparation, advances and application challenges},\ }\href
  {https://doi.org/https://doi.org/10.1016/j.mtcomm.2022.104190} {\bibfield
  {journal} {\bibinfo  {journal} {Materials Today Communications}\ }\textbf
  {\bibinfo {volume} {33}},\ \bibinfo {pages} {104190} (\bibinfo {year}
  {2022})}\BibitemShut {NoStop}%
\bibitem [{\citenamefont {Huang}\ \emph {et~al.}(2023)\citenamefont {Huang},
  \citenamefont {Querales-Flores}, \citenamefont {Teitelbaum}, \citenamefont
  {Cao}, \citenamefont {Henighan}, \citenamefont {Liu}, \citenamefont {Jiang},
  \citenamefont {De~la Pe\~na}, \citenamefont {Krapivin}, \citenamefont
  {Haber}, \citenamefont {Sato}, \citenamefont {Chollet}, \citenamefont {Zhu},
  \citenamefont {Katayama}, \citenamefont {Power}, \citenamefont {Allen},
  \citenamefont {Rotundu}, \citenamefont {Bailey}, \citenamefont {Uher},
  \citenamefont {Trigo}, \citenamefont {Kirchmann}, \citenamefont {Murray},
  \citenamefont {Shen}, \citenamefont {Savi\ifmmode~\acute{c}\else \'{c}\fi{}},
  \citenamefont {Fahy}, \citenamefont {Sobota},\ and\ \citenamefont
  {Reis}}]{Huang2023}%
  \BibitemOpen
  \bibfield  {author} {\bibinfo {author} {\bibfnamefont {Y.}~\bibnamefont
  {Huang}}, \bibinfo {author} {\bibfnamefont {J.~D.}\ \bibnamefont
  {Querales-Flores}}, \bibinfo {author} {\bibfnamefont {S.~W.}\ \bibnamefont
  {Teitelbaum}}, \bibinfo {author} {\bibfnamefont {J.}~\bibnamefont {Cao}},
  \bibinfo {author} {\bibfnamefont {T.}~\bibnamefont {Henighan}}, \bibinfo
  {author} {\bibfnamefont {H.}~\bibnamefont {Liu}}, \bibinfo {author}
  {\bibfnamefont {M.}~\bibnamefont {Jiang}}, \bibinfo {author} {\bibfnamefont
  {G.}~\bibnamefont {De~la Pe\~na}}, \bibinfo {author} {\bibfnamefont
  {V.}~\bibnamefont {Krapivin}}, \bibinfo {author} {\bibfnamefont
  {J.}~\bibnamefont {Haber}}, \bibinfo {author} {\bibfnamefont
  {T.}~\bibnamefont {Sato}}, \bibinfo {author} {\bibfnamefont {M.}~\bibnamefont
  {Chollet}}, \bibinfo {author} {\bibfnamefont {D.}~\bibnamefont {Zhu}},
  \bibinfo {author} {\bibfnamefont {T.}~\bibnamefont {Katayama}}, \bibinfo
  {author} {\bibfnamefont {R.}~\bibnamefont {Power}}, \bibinfo {author}
  {\bibfnamefont {M.}~\bibnamefont {Allen}}, \bibinfo {author} {\bibfnamefont
  {C.~R.}\ \bibnamefont {Rotundu}}, \bibinfo {author} {\bibfnamefont {T.~P.}\
  \bibnamefont {Bailey}}, \bibinfo {author} {\bibfnamefont {C.}~\bibnamefont
  {Uher}}, \bibinfo {author} {\bibfnamefont {M.}~\bibnamefont {Trigo}},
  \bibinfo {author} {\bibfnamefont {P.~S.}\ \bibnamefont {Kirchmann}}, \bibinfo
  {author} {\bibfnamefont {E.~D.}\ \bibnamefont {Murray}}, \bibinfo {author}
  {\bibfnamefont {Z.-X.}\ \bibnamefont {Shen}}, \bibinfo {author}
  {\bibfnamefont {I.}~\bibnamefont {Savi\ifmmode~\acute{c}\else \'{c}\fi{}}},
  \bibinfo {author} {\bibfnamefont {S.}~\bibnamefont {Fahy}}, \bibinfo {author}
  {\bibfnamefont {J.~A.}\ \bibnamefont {Sobota}},\ and\ \bibinfo {author}
  {\bibfnamefont {D.~A.}\ \bibnamefont {Reis}},\ }\bibfield  {title} {\bibinfo
  {title} {Ultrafast measurements of mode-specific deformation potentials of
  ${\mathrm{bi}}_{2}{\mathrm{te}}_{3}$ and
  ${\mathrm{bi}}_{2}{\mathrm{se}}_{3}$},\ }\href
  {https://doi.org/10.1103/PhysRevX.13.041050} {\bibfield  {journal} {\bibinfo
  {journal} {Phys. Rev. X}\ }\textbf {\bibinfo {volume} {13}},\ \bibinfo
  {pages} {041050} (\bibinfo {year} {2023})}\BibitemShut {NoStop}%
\bibitem [{\citenamefont {Cacho}\ \emph {et~al.}(2015)\citenamefont {Cacho},
  \citenamefont {Crepaldi}, \citenamefont {Battiato}, \citenamefont {Braun},
  \citenamefont {Cilento}, \citenamefont {Zacchigna}, \citenamefont {Richter},
  \citenamefont {Heckmann}, \citenamefont {Springate}, \citenamefont {Liu},
  \citenamefont {Dhesi}, \citenamefont {Berger}, \citenamefont {Bugnon},
  \citenamefont {Held}, \citenamefont {Grioni}, \citenamefont {Ebert},
  \citenamefont {Hricovini}, \citenamefont {Min{\'{a}}r},\ and\ \citenamefont
  {Parmigiani}}]{Cacho2015}%
  \BibitemOpen
  \bibfield  {author} {\bibinfo {author} {\bibfnamefont {C.}~\bibnamefont
  {Cacho}}, \bibinfo {author} {\bibfnamefont {A.}~\bibnamefont {Crepaldi}},
  \bibinfo {author} {\bibfnamefont {M.}~\bibnamefont {Battiato}}, \bibinfo
  {author} {\bibfnamefont {J.}~\bibnamefont {Braun}}, \bibinfo {author}
  {\bibfnamefont {F.}~\bibnamefont {Cilento}}, \bibinfo {author} {\bibfnamefont
  {M.}~\bibnamefont {Zacchigna}}, \bibinfo {author} {\bibfnamefont
  {M.}~\bibnamefont {Richter}}, \bibinfo {author} {\bibfnamefont
  {O.}~\bibnamefont {Heckmann}}, \bibinfo {author} {\bibfnamefont
  {E.}~\bibnamefont {Springate}}, \bibinfo {author} {\bibfnamefont
  {Y.}~\bibnamefont {Liu}}, \bibinfo {author} {\bibfnamefont {S.}~\bibnamefont
  {Dhesi}}, \bibinfo {author} {\bibfnamefont {H.}~\bibnamefont {Berger}},
  \bibinfo {author} {\bibfnamefont {P.}~\bibnamefont {Bugnon}}, \bibinfo
  {author} {\bibfnamefont {K.}~\bibnamefont {Held}}, \bibinfo {author}
  {\bibfnamefont {M.}~\bibnamefont {Grioni}}, \bibinfo {author} {\bibfnamefont
  {H.}~\bibnamefont {Ebert}}, \bibinfo {author} {\bibfnamefont
  {K.}~\bibnamefont {Hricovini}}, \bibinfo {author} {\bibfnamefont
  {J.}~\bibnamefont {Min{\'{a}}r}},\ and\ \bibinfo {author} {\bibfnamefont
  {F.}~\bibnamefont {Parmigiani}},\ }\bibfield  {title} {\bibinfo {title}
  {Momentum-resolved spin dynamics of bulk and surface excited states in the
  topological insulator {Bi$_2$Se$_3$}},\ }\href
  {https://doi.org/10.1103/physrevlett.114.097401} {\bibfield  {journal}
  {\bibinfo  {journal} {Physical Review Letters}\ }\textbf {\bibinfo {volume}
  {114}},\ \bibinfo {pages} {097401} (\bibinfo {year} {2015})}\BibitemShut
  {NoStop}%
\bibitem [{\citenamefont {Jozwiak}\ \emph {et~al.}(2016)\citenamefont
  {Jozwiak}, \citenamefont {Sobota}, \citenamefont {Gotlieb}, \citenamefont
  {Kemper}, \citenamefont {Rotundu}, \citenamefont {Birgeneau}, \citenamefont
  {Hussain}, \citenamefont {Lee}, \citenamefont {Shen},\ and\ \citenamefont
  {Lanzara}}]{Jozwiak2016}%
  \BibitemOpen
  \bibfield  {author} {\bibinfo {author} {\bibfnamefont {C.}~\bibnamefont
  {Jozwiak}}, \bibinfo {author} {\bibfnamefont {J.~A.}\ \bibnamefont {Sobota}},
  \bibinfo {author} {\bibfnamefont {K.}~\bibnamefont {Gotlieb}}, \bibinfo
  {author} {\bibfnamefont {A.~F.}\ \bibnamefont {Kemper}}, \bibinfo {author}
  {\bibfnamefont {C.~R.}\ \bibnamefont {Rotundu}}, \bibinfo {author}
  {\bibfnamefont {R.~J.}\ \bibnamefont {Birgeneau}}, \bibinfo {author}
  {\bibfnamefont {Z.}~\bibnamefont {Hussain}}, \bibinfo {author} {\bibfnamefont
  {D.-H.}\ \bibnamefont {Lee}}, \bibinfo {author} {\bibfnamefont {Z.-X.}\
  \bibnamefont {Shen}},\ and\ \bibinfo {author} {\bibfnamefont
  {A.}~\bibnamefont {Lanzara}},\ }\bibfield  {title} {\bibinfo {title}
  {Spin-polarized surface resonances accompanying topological surface state
  formation},\ }\bibfield  {journal} {\bibinfo  {journal} {Nature
  Communications}\ }\textbf {\bibinfo {volume} {7}},\ \href
  {https://doi.org/10.1038/ncomms13143} {10.1038/ncomms13143} (\bibinfo {year}
  {2016})\BibitemShut {NoStop}%
\bibitem [{\citenamefont {Hedayat}\ \emph {et~al.}(2021)\citenamefont
  {Hedayat}, \citenamefont {Bugini}, \citenamefont {Yi}, \citenamefont {Chen},
  \citenamefont {Zhou}, \citenamefont {Cerullo}, \citenamefont {Dallera},\ and\
  \citenamefont {Carpene}}]{Hedayat2021}%
  \BibitemOpen
  \bibfield  {author} {\bibinfo {author} {\bibfnamefont {H.}~\bibnamefont
  {Hedayat}}, \bibinfo {author} {\bibfnamefont {D.}~\bibnamefont {Bugini}},
  \bibinfo {author} {\bibfnamefont {H.}~\bibnamefont {Yi}}, \bibinfo {author}
  {\bibfnamefont {C.}~\bibnamefont {Chen}}, \bibinfo {author} {\bibfnamefont
  {X.}~\bibnamefont {Zhou}}, \bibinfo {author} {\bibfnamefont {G.}~\bibnamefont
  {Cerullo}}, \bibinfo {author} {\bibfnamefont {C.}~\bibnamefont {Dallera}},\
  and\ \bibinfo {author} {\bibfnamefont {E.}~\bibnamefont {Carpene}},\
  }\bibfield  {title} {\bibinfo {title} {Ultrafast evolution of bulk, surface
  and surface resonance states in photoexcited {Bi$_2$Te$_3$}},\ }\bibfield
  {journal} {\bibinfo  {journal} {Scientific Reports}\ }\textbf {\bibinfo
  {volume} {11}},\ \href {https://doi.org/10.1038/s41598-021-83848-z}
  {10.1038/s41598-021-83848-z} (\bibinfo {year} {2021})\BibitemShut {NoStop}%
\bibitem [{\citenamefont {Bianchi}\ \emph {et~al.}(2010)\citenamefont
  {Bianchi}, \citenamefont {Guan}, \citenamefont {Bao}, \citenamefont {Mi},
  \citenamefont {Iversen}, \citenamefont {King},\ and\ \citenamefont
  {Hofmann}}]{Bianchi2010}%
  \BibitemOpen
  \bibfield  {author} {\bibinfo {author} {\bibfnamefont {M.}~\bibnamefont
  {Bianchi}}, \bibinfo {author} {\bibfnamefont {D.}~\bibnamefont {Guan}},
  \bibinfo {author} {\bibfnamefont {S.}~\bibnamefont {Bao}}, \bibinfo {author}
  {\bibfnamefont {J.}~\bibnamefont {Mi}}, \bibinfo {author} {\bibfnamefont
  {B.~B.}\ \bibnamefont {Iversen}}, \bibinfo {author} {\bibfnamefont {P.~D.}\
  \bibnamefont {King}},\ and\ \bibinfo {author} {\bibfnamefont
  {P.}~\bibnamefont {Hofmann}},\ }\bibfield  {title} {\bibinfo {title}
  {Coexistence of the topological state and a two-dimensional electron gas on
  the surface of {Bi$_2$Se$_3$}},\ }\bibfield  {journal} {\bibinfo  {journal}
  {Nature Communications}\ }\textbf {\bibinfo {volume} {1}},\ \href
  {https://doi.org/10.1038/ncomms1131} {10.1038/ncomms1131} (\bibinfo {year}
  {2010})\BibitemShut {NoStop}%
\bibitem [{\citenamefont {Chen}\ \emph {et~al.}(2012)\citenamefont {Chen},
  \citenamefont {He}, \citenamefont {Weng}, \citenamefont {Zhang},
  \citenamefont {Zhao}, \citenamefont {Liu}, \citenamefont {Jia}, \citenamefont
  {Mou}, \citenamefont {Liu}, \citenamefont {He}, \citenamefont {Peng},
  \citenamefont {Feng}, \citenamefont {Xie}, \citenamefont {Liu}, \citenamefont
  {Dong}, \citenamefont {Zhang}, \citenamefont {Wang}, \citenamefont {Peng},
  \citenamefont {Wang}, \citenamefont {Zhang}, \citenamefont {Yang},
  \citenamefont {Chen}, \citenamefont {Xu}, \citenamefont {Dai}, \citenamefont
  {Fang},\ and\ \citenamefont {Zhou}}]{Chen2012}%
  \BibitemOpen
  \bibfield  {author} {\bibinfo {author} {\bibfnamefont {C.}~\bibnamefont
  {Chen}}, \bibinfo {author} {\bibfnamefont {S.}~\bibnamefont {He}}, \bibinfo
  {author} {\bibfnamefont {H.}~\bibnamefont {Weng}}, \bibinfo {author}
  {\bibfnamefont {W.}~\bibnamefont {Zhang}}, \bibinfo {author} {\bibfnamefont
  {L.}~\bibnamefont {Zhao}}, \bibinfo {author} {\bibfnamefont {H.}~\bibnamefont
  {Liu}}, \bibinfo {author} {\bibfnamefont {X.}~\bibnamefont {Jia}}, \bibinfo
  {author} {\bibfnamefont {D.}~\bibnamefont {Mou}}, \bibinfo {author}
  {\bibfnamefont {S.}~\bibnamefont {Liu}}, \bibinfo {author} {\bibfnamefont
  {J.}~\bibnamefont {He}}, \bibinfo {author} {\bibfnamefont {Y.}~\bibnamefont
  {Peng}}, \bibinfo {author} {\bibfnamefont {Y.}~\bibnamefont {Feng}}, \bibinfo
  {author} {\bibfnamefont {Z.}~\bibnamefont {Xie}}, \bibinfo {author}
  {\bibfnamefont {G.}~\bibnamefont {Liu}}, \bibinfo {author} {\bibfnamefont
  {X.}~\bibnamefont {Dong}}, \bibinfo {author} {\bibfnamefont {J.}~\bibnamefont
  {Zhang}}, \bibinfo {author} {\bibfnamefont {X.}~\bibnamefont {Wang}},
  \bibinfo {author} {\bibfnamefont {Q.}~\bibnamefont {Peng}}, \bibinfo {author}
  {\bibfnamefont {Z.}~\bibnamefont {Wang}}, \bibinfo {author} {\bibfnamefont
  {S.}~\bibnamefont {Zhang}}, \bibinfo {author} {\bibfnamefont
  {F.}~\bibnamefont {Yang}}, \bibinfo {author} {\bibfnamefont {C.}~\bibnamefont
  {Chen}}, \bibinfo {author} {\bibfnamefont {Z.}~\bibnamefont {Xu}}, \bibinfo
  {author} {\bibfnamefont {X.}~\bibnamefont {Dai}}, \bibinfo {author}
  {\bibfnamefont {Z.}~\bibnamefont {Fang}},\ and\ \bibinfo {author}
  {\bibfnamefont {X.~J.}\ \bibnamefont {Zhou}},\ }\bibfield  {title} {\bibinfo
  {title} {Robustness of topological order and formation of quantum well states
  in topological insulators exposed to ambient environment},\ }\href
  {https://doi.org/10.1073/pnas.1115555109} {\bibfield  {journal} {\bibinfo
  {journal} {Proceedings of the National Academy of Sciences}\ }\textbf
  {\bibinfo {volume} {109}},\ \bibinfo {pages} {3694} (\bibinfo {year}
  {2012})}\BibitemShut {NoStop}%
\bibitem [{\citenamefont {Kuroda}\ \emph {et~al.}(2010)\citenamefont {Kuroda},
  \citenamefont {Arita}, \citenamefont {Miyamoto}, \citenamefont {Ye},
  \citenamefont {Jiang}, \citenamefont {Kimura}, \citenamefont {Krasovskii},
  \citenamefont {Chulkov}, \citenamefont {Iwasawa}, \citenamefont {Okuda},
  \citenamefont {Shimada}, \citenamefont {Ueda}, \citenamefont {Namatame},\
  and\ \citenamefont {Taniguchi}}]{Kuroda2010}%
  \BibitemOpen
  \bibfield  {author} {\bibinfo {author} {\bibfnamefont {K.}~\bibnamefont
  {Kuroda}}, \bibinfo {author} {\bibfnamefont {M.}~\bibnamefont {Arita}},
  \bibinfo {author} {\bibfnamefont {K.}~\bibnamefont {Miyamoto}}, \bibinfo
  {author} {\bibfnamefont {M.}~\bibnamefont {Ye}}, \bibinfo {author}
  {\bibfnamefont {J.}~\bibnamefont {Jiang}}, \bibinfo {author} {\bibfnamefont
  {A.}~\bibnamefont {Kimura}}, \bibinfo {author} {\bibfnamefont {E.~E.}\
  \bibnamefont {Krasovskii}}, \bibinfo {author} {\bibfnamefont {E.~V.}\
  \bibnamefont {Chulkov}}, \bibinfo {author} {\bibfnamefont {H.}~\bibnamefont
  {Iwasawa}}, \bibinfo {author} {\bibfnamefont {T.}~\bibnamefont {Okuda}},
  \bibinfo {author} {\bibfnamefont {K.}~\bibnamefont {Shimada}}, \bibinfo
  {author} {\bibfnamefont {Y.}~\bibnamefont {Ueda}}, \bibinfo {author}
  {\bibfnamefont {H.}~\bibnamefont {Namatame}},\ and\ \bibinfo {author}
  {\bibfnamefont {M.}~\bibnamefont {Taniguchi}},\ }\bibfield  {title} {\bibinfo
  {title} {Hexagonally deformed {Fermi} surface of the {3D} topological
  insulator {Bi$_2$Se$_3$}},\ }\href
  {https://doi.org/10.1103/physrevlett.105.076802} {\bibfield  {journal}
  {\bibinfo  {journal} {Physical Review Letters}\ }\textbf {\bibinfo {volume}
  {105}},\ \bibinfo {pages} {076802} (\bibinfo {year} {2010})}\BibitemShut
  {NoStop}%
\bibitem [{\citenamefont {S{\'{a}}nchez-Barriga}\ \emph
  {et~al.}(2014)\citenamefont {S{\'{a}}nchez-Barriga}, \citenamefont {Scholz},
  \citenamefont {Golias}, \citenamefont {Rienks}, \citenamefont {Marchenko},
  \citenamefont {Varykhalov}, \citenamefont {Yashina},\ and\ \citenamefont
  {Rader}}]{SanchezBarriga2014}%
  \BibitemOpen
  \bibfield  {author} {\bibinfo {author} {\bibfnamefont {J.}~\bibnamefont
  {S{\'{a}}nchez-Barriga}}, \bibinfo {author} {\bibfnamefont {M.~R.}\
  \bibnamefont {Scholz}}, \bibinfo {author} {\bibfnamefont {E.}~\bibnamefont
  {Golias}}, \bibinfo {author} {\bibfnamefont {E.}~\bibnamefont {Rienks}},
  \bibinfo {author} {\bibfnamefont {D.}~\bibnamefont {Marchenko}}, \bibinfo
  {author} {\bibfnamefont {A.}~\bibnamefont {Varykhalov}}, \bibinfo {author}
  {\bibfnamefont {L.~V.}\ \bibnamefont {Yashina}},\ and\ \bibinfo {author}
  {\bibfnamefont {O.}~\bibnamefont {Rader}},\ }\bibfield  {title} {\bibinfo
  {title} {{Anisotropic effect of warping on the lifetime broadening of
  topological surface states in angle-resolved photoemission from
  {$\mathrm{Bi_2Te_3}$}}},\ }\href {https://doi.org/10.1103/physrevb.90.195413}
  {\bibfield  {journal} {\bibinfo  {journal} {Physical Review B}\ }\textbf
  {\bibinfo {volume} {90}},\ \bibinfo {pages} {195413} (\bibinfo {year}
  {2014})}\BibitemShut {NoStop}%
\bibitem [{\citenamefont {Nomura}\ \emph {et~al.}(2014)\citenamefont {Nomura},
  \citenamefont {Souma}, \citenamefont {Takayama}, \citenamefont {Sato},
  \citenamefont {Takahashi}, \citenamefont {Eto}, \citenamefont {Segawa},\ and\
  \citenamefont {Ando}}]{Nomura2014}%
  \BibitemOpen
  \bibfield  {author} {\bibinfo {author} {\bibfnamefont {M.}~\bibnamefont
  {Nomura}}, \bibinfo {author} {\bibfnamefont {S.}~\bibnamefont {Souma}},
  \bibinfo {author} {\bibfnamefont {A.}~\bibnamefont {Takayama}}, \bibinfo
  {author} {\bibfnamefont {T.}~\bibnamefont {Sato}}, \bibinfo {author}
  {\bibfnamefont {T.}~\bibnamefont {Takahashi}}, \bibinfo {author}
  {\bibfnamefont {K.}~\bibnamefont {Eto}}, \bibinfo {author} {\bibfnamefont
  {K.}~\bibnamefont {Segawa}},\ and\ \bibinfo {author} {\bibfnamefont
  {Y.}~\bibnamefont {Ando}},\ }\bibfield  {title} {\bibinfo {title}
  {Relationship between {Fermi} surface warping and out-of-plane spin
  polarization in topological insulators: A view from spin- and angle-resolved
  photoemission},\ }\href {https://doi.org/10.1103/physrevb.89.045134}
  {\bibfield  {journal} {\bibinfo  {journal} {Physical Review B}\ }\textbf
  {\bibinfo {volume} {89}},\ \bibinfo {pages} {045134} (\bibinfo {year}
  {2014})}\BibitemShut {NoStop}%
\bibitem [{\citenamefont {Nurmamat}\ \emph {et~al.}(2018)\citenamefont
  {Nurmamat}, \citenamefont {Krasovskii}, \citenamefont {Ishida}, \citenamefont
  {Sumida}, \citenamefont {Chen}, \citenamefont {Yoshikawa}, \citenamefont
  {Chulkov}, \citenamefont {Kokh}, \citenamefont {Tereshchenko}, \citenamefont
  {Shin},\ and\ \citenamefont {Kimura}}]{Nurmamat2018}%
  \BibitemOpen
  \bibfield  {author} {\bibinfo {author} {\bibfnamefont {M.}~\bibnamefont
  {Nurmamat}}, \bibinfo {author} {\bibfnamefont {E.~E.}\ \bibnamefont
  {Krasovskii}}, \bibinfo {author} {\bibfnamefont {Y.}~\bibnamefont {Ishida}},
  \bibinfo {author} {\bibfnamefont {K.}~\bibnamefont {Sumida}}, \bibinfo
  {author} {\bibfnamefont {J.}~\bibnamefont {Chen}}, \bibinfo {author}
  {\bibfnamefont {T.}~\bibnamefont {Yoshikawa}}, \bibinfo {author}
  {\bibfnamefont {E.~V.}\ \bibnamefont {Chulkov}}, \bibinfo {author}
  {\bibfnamefont {K.~A.}\ \bibnamefont {Kokh}}, \bibinfo {author}
  {\bibfnamefont {O.~E.}\ \bibnamefont {Tereshchenko}}, \bibinfo {author}
  {\bibfnamefont {S.}~\bibnamefont {Shin}},\ and\ \bibinfo {author}
  {\bibfnamefont {A.}~\bibnamefont {Kimura}},\ }\bibfield  {title} {\bibinfo
  {title} {Ultrafast dynamics of an unoccupied surface resonance state in
  {BiTe$_2$Se}},\ }\href {https://doi.org/10.1103/physrevb.97.115303}
  {\bibfield  {journal} {\bibinfo  {journal} {Physical Review B}\ }\textbf
  {\bibinfo {volume} {97}},\ \bibinfo {pages} {115303} (\bibinfo {year}
  {2018})}\BibitemShut {NoStop}%
\bibitem [{\citenamefont {Basak}\ \emph {et~al.}(2011)\citenamefont {Basak},
  \citenamefont {Lin}, \citenamefont {Wray}, \citenamefont {Xu}, \citenamefont
  {Fu}, \citenamefont {Hasan},\ and\ \citenamefont {Bansil}}]{Basak2011}%
  \BibitemOpen
  \bibfield  {author} {\bibinfo {author} {\bibfnamefont {S.}~\bibnamefont
  {Basak}}, \bibinfo {author} {\bibfnamefont {H.}~\bibnamefont {Lin}}, \bibinfo
  {author} {\bibfnamefont {L.~A.}\ \bibnamefont {Wray}}, \bibinfo {author}
  {\bibfnamefont {S.-Y.}\ \bibnamefont {Xu}}, \bibinfo {author} {\bibfnamefont
  {L.}~\bibnamefont {Fu}}, \bibinfo {author} {\bibfnamefont {M.~Z.}\
  \bibnamefont {Hasan}},\ and\ \bibinfo {author} {\bibfnamefont
  {A.}~\bibnamefont {Bansil}},\ }\bibfield  {title} {\bibinfo {title} {Spin
  texture on the warped {Dirac}-cone surface states in topological
  insulators},\ }\href {https://doi.org/10.1103/physrevb.84.121401} {\bibfield
  {journal} {\bibinfo  {journal} {Physical Review B}\ }\textbf {\bibinfo
  {volume} {84}},\ \bibinfo {pages} {121401} (\bibinfo {year}
  {2011})}\BibitemShut {NoStop}%
\bibitem [{\citenamefont {Wang}\ \emph {et~al.}(2011)\citenamefont {Wang},
  \citenamefont {Hsieh}, \citenamefont {Pilon}, \citenamefont {Fu},
  \citenamefont {Gardner}, \citenamefont {Lee},\ and\ \citenamefont
  {Gedik}}]{Wang2011}%
  \BibitemOpen
  \bibfield  {author} {\bibinfo {author} {\bibfnamefont {Y.~H.}\ \bibnamefont
  {Wang}}, \bibinfo {author} {\bibfnamefont {D.}~\bibnamefont {Hsieh}},
  \bibinfo {author} {\bibfnamefont {D.}~\bibnamefont {Pilon}}, \bibinfo
  {author} {\bibfnamefont {L.}~\bibnamefont {Fu}}, \bibinfo {author}
  {\bibfnamefont {D.~R.}\ \bibnamefont {Gardner}}, \bibinfo {author}
  {\bibfnamefont {Y.~S.}\ \bibnamefont {Lee}},\ and\ \bibinfo {author}
  {\bibfnamefont {N.}~\bibnamefont {Gedik}},\ }\bibfield  {title} {\bibinfo
  {title} {Observation of a warped helical spin texture in {Bi$_2$Se$_3$} from
  circular dichroism angle-resolved photoemission spectroscopy},\ }\href
  {https://doi.org/10.1103/physrevlett.107.207602} {\bibfield  {journal}
  {\bibinfo  {journal} {Physical Review Letters}\ }\textbf {\bibinfo {volume}
  {107}},\ \bibinfo {pages} {207602} (\bibinfo {year} {2011})}\BibitemShut
  {NoStop}%
\bibitem [{\citenamefont {Fu}(2009)}]{Fu2009}%
  \BibitemOpen
  \bibfield  {author} {\bibinfo {author} {\bibfnamefont {L.}~\bibnamefont
  {Fu}},\ }\bibfield  {title} {\bibinfo {title} {Hexagonal warping effects in
  the surface states of the topological insulator {Bi$_2$Te$_3$}},\ }\href
  {https://doi.org/10.1103/physrevlett.103.266801} {\bibfield  {journal}
  {\bibinfo  {journal} {Physical Review Letters}\ }\textbf {\bibinfo {volume}
  {103}},\ \bibinfo {pages} {266801} (\bibinfo {year} {2009})}\BibitemShut
  {NoStop}%
\bibitem [{\citenamefont {Liu}\ \emph {et~al.}(2010)\citenamefont {Liu},
  \citenamefont {Qi}, \citenamefont {Zhang}, \citenamefont {Dai}, \citenamefont
  {Fang},\ and\ \citenamefont {Zhang}}]{Liu2010}%
  \BibitemOpen
  \bibfield  {author} {\bibinfo {author} {\bibfnamefont {C.-X.}\ \bibnamefont
  {Liu}}, \bibinfo {author} {\bibfnamefont {X.-L.}\ \bibnamefont {Qi}},
  \bibinfo {author} {\bibfnamefont {H.}~\bibnamefont {Zhang}}, \bibinfo
  {author} {\bibfnamefont {X.}~\bibnamefont {Dai}}, \bibinfo {author}
  {\bibfnamefont {Z.}~\bibnamefont {Fang}},\ and\ \bibinfo {author}
  {\bibfnamefont {S.-C.}\ \bibnamefont {Zhang}},\ }\bibfield  {title} {\bibinfo
  {title} {Model hamiltonian for topological insulators},\ }\href
  {https://doi.org/10.1103/physrevb.82.045122} {\bibfield  {journal} {\bibinfo
  {journal} {Physical Review B}\ }\textbf {\bibinfo {volume} {82}},\ \bibinfo
  {pages} {045122} (\bibinfo {year} {2010})}\BibitemShut {NoStop}%
\bibitem [{\citenamefont {Nuber}\ \emph {et~al.}(2011)\citenamefont {Nuber},
  \citenamefont {Braun}, \citenamefont {Forster}, \citenamefont {Min{\'{a}}r},
  \citenamefont {Reinert},\ and\ \citenamefont {Ebert}}]{Nuber2011}%
  \BibitemOpen
  \bibfield  {author} {\bibinfo {author} {\bibfnamefont {A.}~\bibnamefont
  {Nuber}}, \bibinfo {author} {\bibfnamefont {J.}~\bibnamefont {Braun}},
  \bibinfo {author} {\bibfnamefont {F.}~\bibnamefont {Forster}}, \bibinfo
  {author} {\bibfnamefont {J.}~\bibnamefont {Min{\'{a}}r}}, \bibinfo {author}
  {\bibfnamefont {F.}~\bibnamefont {Reinert}},\ and\ \bibinfo {author}
  {\bibfnamefont {H.}~\bibnamefont {Ebert}},\ }\bibfield  {title} {\bibinfo
  {title} {Surface versus bulk contributions to the {Rashba} splitting in
  surface systems},\ }\href {https://doi.org/10.1103/physrevb.83.165401}
  {\bibfield  {journal} {\bibinfo  {journal} {Physical Review B}\ }\textbf
  {\bibinfo {volume} {83}},\ \bibinfo {pages} {165401} (\bibinfo {year}
  {2011})}\BibitemShut {NoStop}%
\bibitem [{\citenamefont {Jain}\ \emph {et~al.}(2013)\citenamefont {Jain},
  \citenamefont {Ong}, \citenamefont {Hautier}, \citenamefont {Chen},
  \citenamefont {Richards}, \citenamefont {Dacek}, \citenamefont {Cholia},
  \citenamefont {Gunter}, \citenamefont {Skinner}, \citenamefont {Ceder},\ and\
  \citenamefont {Persson}}]{Jain2013}%
  \BibitemOpen
  \bibfield  {author} {\bibinfo {author} {\bibfnamefont {A.}~\bibnamefont
  {Jain}}, \bibinfo {author} {\bibfnamefont {S.~P.}\ \bibnamefont {Ong}},
  \bibinfo {author} {\bibfnamefont {G.}~\bibnamefont {Hautier}}, \bibinfo
  {author} {\bibfnamefont {W.}~\bibnamefont {Chen}}, \bibinfo {author}
  {\bibfnamefont {W.~D.}\ \bibnamefont {Richards}}, \bibinfo {author}
  {\bibfnamefont {S.}~\bibnamefont {Dacek}}, \bibinfo {author} {\bibfnamefont
  {S.}~\bibnamefont {Cholia}}, \bibinfo {author} {\bibfnamefont
  {D.}~\bibnamefont {Gunter}}, \bibinfo {author} {\bibfnamefont
  {D.}~\bibnamefont {Skinner}}, \bibinfo {author} {\bibfnamefont
  {G.}~\bibnamefont {Ceder}},\ and\ \bibinfo {author} {\bibfnamefont {K.~A.}\
  \bibnamefont {Persson}},\ }\bibfield  {title} {\bibinfo {title} {Commentary:
  {The Materials Project}: A materials genome approach to accelerating
  materials innovation},\ }\bibfield  {journal} {\bibinfo  {journal} {APL
  Materials}\ }\textbf {\bibinfo {volume} {1}},\ \href
  {https://doi.org/10.1063/1.4812323} {10.1063/1.4812323} (\bibinfo {year}
  {2013})\BibitemShut {NoStop}%
\bibitem [{Mat()}]{MatPrj}%
  \BibitemOpen
  \href {http://www.materialsproject.org/} {\bibinfo {title} {{The Materials
  project}}},\ \bibinfo {note} {accessed: February 2024}\BibitemShut {NoStop}%
\bibitem [{\citenamefont {Momma}\ and\ \citenamefont
  {Izumi}(2008)}]{Momma2008}%
  \BibitemOpen
  \bibfield  {author} {\bibinfo {author} {\bibfnamefont {K.}~\bibnamefont
  {Momma}}\ and\ \bibinfo {author} {\bibfnamefont {F.}~\bibnamefont {Izumi}},\
  }\bibfield  {title} {\bibinfo {title} {{\emph{VESTA}}: a three-dimensional
  visualization system for electronic and structural analysis},\ }\href
  {https://doi.org/10.1107/s0021889808012016} {\bibfield  {journal} {\bibinfo
  {journal} {Journal of Applied Crystallography}\ }\textbf {\bibinfo {volume}
  {41}},\ \bibinfo {pages} {653} (\bibinfo {year} {2008})}\BibitemShut
  {NoStop}%
\bibitem [{\citenamefont {Malmstr{\"o}m}\ and\ \citenamefont
  {Rundgren}(1980)}]{Malmstroem1980}%
  \BibitemOpen
  \bibfield  {author} {\bibinfo {author} {\bibfnamefont {G.}~\bibnamefont
  {Malmstr{\"o}m}}\ and\ \bibinfo {author} {\bibfnamefont {J.}~\bibnamefont
  {Rundgren}},\ }\bibfield  {title} {\bibinfo {title} {A program for
  calculation of the reflection and transmission of electrons through a surface
  potential barrier},\ }\href {https://doi.org/10.1016/0010-4655(80)90053-3}
  {\bibfield  {journal} {\bibinfo  {journal} {Computer Physics Communications}\
  }\textbf {\bibinfo {volume} {19}},\ \bibinfo {pages} {263} (\bibinfo {year}
  {1980})}\BibitemShut {NoStop}%
\bibitem [{\citenamefont {Braun}\ \emph {et~al.}(2018)\citenamefont {Braun},
  \citenamefont {Min{\'{a}}r},\ and\ \citenamefont {Ebert}}]{Braun2018}%
  \BibitemOpen
  \bibfield  {author} {\bibinfo {author} {\bibfnamefont {J.}~\bibnamefont
  {Braun}}, \bibinfo {author} {\bibfnamefont {J.}~\bibnamefont {Min{\'{a}}r}},\
  and\ \bibinfo {author} {\bibfnamefont {H.}~\bibnamefont {Ebert}},\ }\bibfield
   {title} {\bibinfo {title} {Correlation, temperature and disorder: Recent
  developments in the one-step description of angle-resolved photoemission},\
  }\href {https://doi.org/10.1016/j.physrep.2018.02.007} {\bibfield  {journal}
  {\bibinfo  {journal} {Physics Reports}\ }\textbf {\bibinfo {volume} {740}},\
  \bibinfo {pages} {1} (\bibinfo {year} {2018})}\BibitemShut {NoStop}%
\bibitem [{\citenamefont {Bendounan}\ \emph {et~al.}(2011)\citenamefont
  {Bendounan}, \citenamefont {A\"it-Mansour}, \citenamefont {Braun},
  \citenamefont {Min{\'{a}}r}, \citenamefont {Bornemann}, \citenamefont
  {Fasel}, \citenamefont {Gr\"oning}, \citenamefont {Sirotti},\ and\
  \citenamefont {Ebert}}]{Bendounan2011}%
  \BibitemOpen
  \bibfield  {author} {\bibinfo {author} {\bibfnamefont {A.}~\bibnamefont
  {Bendounan}}, \bibinfo {author} {\bibfnamefont {K.}~\bibnamefont
  {A\"it-Mansour}}, \bibinfo {author} {\bibfnamefont {J.}~\bibnamefont
  {Braun}}, \bibinfo {author} {\bibfnamefont {J.}~\bibnamefont {Min{\'{a}}r}},
  \bibinfo {author} {\bibfnamefont {S.}~\bibnamefont {Bornemann}}, \bibinfo
  {author} {\bibfnamefont {R.}~\bibnamefont {Fasel}}, \bibinfo {author}
  {\bibfnamefont {O.}~\bibnamefont {Gr\"oning}}, \bibinfo {author}
  {\bibfnamefont {F.}~\bibnamefont {Sirotti}},\ and\ \bibinfo {author}
  {\bibfnamefont {H.}~\bibnamefont {Ebert}},\ }\bibfield  {title} {\bibinfo
  {title} {Evolution of the {Rashba} spin-orbit-split {Shockley} state on
  {Ag/Pt(111)}},\ }\href {https://doi.org/10.1103/physrevb.83.195427}
  {\bibfield  {journal} {\bibinfo  {journal} {Physical Review B}\ }\textbf
  {\bibinfo {volume} {83}},\ \bibinfo {pages} {195427} (\bibinfo {year}
  {2011})}\BibitemShut {NoStop}%
\bibitem [{\citenamefont {Ebert}()}]{Ebe22}%
  \BibitemOpen
  \bibfield  {author} {\bibinfo {author} {\bibfnamefont {H.}~\bibnamefont
  {Ebert}},\ }\href
  {https://www.ebert.cup.uni-muenchen.de/en/software-en/13-sprkkr} {\bibinfo
  {title} {et al.\emph{,} {The Munich SPR-KKR package, version 8.6}}},\
  \bibinfo {note} {accessed: February 2024}\BibitemShut {NoStop}%
\bibitem [{\citenamefont {Perdew}\ \emph {et~al.}(1996)\citenamefont {Perdew},
  \citenamefont {Burke},\ and\ \citenamefont {Ernzerhof}}]{Perdew1996}%
  \BibitemOpen
  \bibfield  {author} {\bibinfo {author} {\bibfnamefont {J.~P.}\ \bibnamefont
  {Perdew}}, \bibinfo {author} {\bibfnamefont {K.}~\bibnamefont {Burke}},\ and\
  \bibinfo {author} {\bibfnamefont {M.}~\bibnamefont {Ernzerhof}},\ }\bibfield
  {title} {\bibinfo {title} {Generalized gradient approximation made simple},\
  }\href {https://doi.org/10.1103/physrevlett.77.3865} {\bibfield  {journal}
  {\bibinfo  {journal} {Physical Review Letters}\ }\textbf {\bibinfo {volume}
  {77}},\ \bibinfo {pages} {3865} (\bibinfo {year} {1996})}\BibitemShut
  {NoStop}%
\bibitem [{\citenamefont {Braun}\ \emph {et~al.}(2014)\citenamefont {Braun},
  \citenamefont {Miyamoto}, \citenamefont {Kimura}, \citenamefont {Okuda},
  \citenamefont {Donath}, \citenamefont {Ebert},\ and\ \citenamefont
  {Min{\'{a}}r}}]{Braun2014}%
  \BibitemOpen
  \bibfield  {author} {\bibinfo {author} {\bibfnamefont {J.}~\bibnamefont
  {Braun}}, \bibinfo {author} {\bibfnamefont {K.}~\bibnamefont {Miyamoto}},
  \bibinfo {author} {\bibfnamefont {A.}~\bibnamefont {Kimura}}, \bibinfo
  {author} {\bibfnamefont {T.}~\bibnamefont {Okuda}}, \bibinfo {author}
  {\bibfnamefont {M.}~\bibnamefont {Donath}}, \bibinfo {author} {\bibfnamefont
  {H.}~\bibnamefont {Ebert}},\ and\ \bibinfo {author} {\bibfnamefont
  {J.}~\bibnamefont {Min{\'{a}}r}},\ }\bibfield  {title} {\bibinfo {title}
  {Exceptional behavior of d-like surface resonances on {{W(110)}}: the
  one-step model in its density matrix formulation},\ }\href
  {https://doi.org/10.1088/1367-2630/16/1/015005} {\bibfield  {journal}
  {\bibinfo  {journal} {New Journal of Physics}\ }\textbf {\bibinfo {volume}
  {16}},\ \bibinfo {pages} {015005} (\bibinfo {year} {2014})}\BibitemShut
  {NoStop}%
\bibitem [{\citenamefont {Grass}\ \emph {et~al.}(1993)\citenamefont {Grass},
  \citenamefont {Braun}, \citenamefont {Borstel}, \citenamefont {Schneider},
  \citenamefont {Durr}, \citenamefont {Fauster},\ and\ \citenamefont
  {Dose}}]{Grass1993}%
  \BibitemOpen
  \bibfield  {author} {\bibinfo {author} {\bibfnamefont {M.}~\bibnamefont
  {Grass}}, \bibinfo {author} {\bibfnamefont {J.}~\bibnamefont {Braun}},
  \bibinfo {author} {\bibfnamefont {G.}~\bibnamefont {Borstel}}, \bibinfo
  {author} {\bibfnamefont {R.}~\bibnamefont {Schneider}}, \bibinfo {author}
  {\bibfnamefont {H.}~\bibnamefont {Durr}}, \bibinfo {author} {\bibfnamefont
  {T.}~\bibnamefont {Fauster}},\ and\ \bibinfo {author} {\bibfnamefont
  {V.}~\bibnamefont {Dose}},\ }\bibfield  {title} {\bibinfo {title} {Unoccupied
  electronic states and surface barriers at {Cu} surfaces},\ }\href
  {https://doi.org/10.1088/0953-8984/5/5/011} {\bibfield  {journal} {\bibinfo
  {journal} {Journal of Physics: Condensed Matter}\ }\textbf {\bibinfo {volume}
  {5}},\ \bibinfo {pages} {599} (\bibinfo {year} {1993})}\BibitemShut {NoStop}%
\bibitem [{\citenamefont {Datzer}\ \emph {et~al.}(2017)\citenamefont {Datzer},
  \citenamefont {Zumb{\"u}lte}, \citenamefont {Braun}, \citenamefont
  {F{\"o}rster}, \citenamefont {Schmidt}, \citenamefont {Mi}, \citenamefont
  {Iversen}, \citenamefont {Hofmann}, \citenamefont {Min{\'{a}}r},
  \citenamefont {Ebert}, \citenamefont {Kr{\"u}ger}, \citenamefont {Rohlfing},\
  and\ \citenamefont {Donath}}]{Datzer2017}%
  \BibitemOpen
  \bibfield  {author} {\bibinfo {author} {\bibfnamefont {C.}~\bibnamefont
  {Datzer}}, \bibinfo {author} {\bibfnamefont {A.}~\bibnamefont
  {Zumb{\"u}lte}}, \bibinfo {author} {\bibfnamefont {J.}~\bibnamefont {Braun}},
  \bibinfo {author} {\bibfnamefont {T.}~\bibnamefont {F{\"o}rster}}, \bibinfo
  {author} {\bibfnamefont {A.~B.}\ \bibnamefont {Schmidt}}, \bibinfo {author}
  {\bibfnamefont {J.}~\bibnamefont {Mi}}, \bibinfo {author} {\bibfnamefont
  {B.}~\bibnamefont {Iversen}}, \bibinfo {author} {\bibfnamefont
  {P.}~\bibnamefont {Hofmann}}, \bibinfo {author} {\bibfnamefont
  {J.}~\bibnamefont {Min{\'{a}}r}}, \bibinfo {author} {\bibfnamefont
  {H.}~\bibnamefont {Ebert}}, \bibinfo {author} {\bibfnamefont
  {P.}~\bibnamefont {Kr{\"u}ger}}, \bibinfo {author} {\bibfnamefont
  {M.}~\bibnamefont {Rohlfing}},\ and\ \bibinfo {author} {\bibfnamefont
  {M.}~\bibnamefont {Donath}},\ }\bibfield  {title} {\bibinfo {title}
  {Unraveling the spin structure of unoccupied states in {Bi$_2$Se$_3$}},\
  }\href {https://doi.org/10.1103/physrevb.95.115401} {\bibfield  {journal}
  {\bibinfo  {journal} {Physical Review B}\ }\textbf {\bibinfo {volume} {95}},\
  \bibinfo {pages} {115401} (\bibinfo {year} {2017})}\BibitemShut {NoStop}%
\bibitem [{\citenamefont {Peli}\ \emph {et~al.}(2020)\citenamefont {Peli},
  \citenamefont {Puntel}, \citenamefont {Kopic}, \citenamefont {Sockol},
  \citenamefont {Parmigiani},\ and\ \citenamefont {Cilento}}]{Peli2020}%
  \BibitemOpen
  \bibfield  {author} {\bibinfo {author} {\bibfnamefont {S.}~\bibnamefont
  {Peli}}, \bibinfo {author} {\bibfnamefont {D.}~\bibnamefont {Puntel}},
  \bibinfo {author} {\bibfnamefont {D.}~\bibnamefont {Kopic}}, \bibinfo
  {author} {\bibfnamefont {B.}~\bibnamefont {Sockol}}, \bibinfo {author}
  {\bibfnamefont {F.}~\bibnamefont {Parmigiani}},\ and\ \bibinfo {author}
  {\bibfnamefont {F.}~\bibnamefont {Cilento}},\ }\bibfield  {title} {\bibinfo
  {title} {Time-resolved {VUV} {ARPES} at 10.8 {eV} photon energy and {MHz}
  repetition rate},\ }\href {https://doi.org/10.1016/j.elspec.2020.146978}
  {\bibfield  {journal} {\bibinfo  {journal} {Journal of Electron Spectroscopy
  and Related Phenomena}\ }\textbf {\bibinfo {volume} {243}},\ \bibinfo {pages}
  {146978} (\bibinfo {year} {2020})}\BibitemShut {NoStop}%
\bibitem [{\citenamefont {Crepaldi}\ \emph {et~al.}(2012)\citenamefont
  {Crepaldi}, \citenamefont {Ressel}, \citenamefont {Cilento}, \citenamefont
  {Zacchigna}, \citenamefont {Grazioli}, \citenamefont {Berger}, \citenamefont
  {Bugnon}, \citenamefont {Kern}, \citenamefont {Grioni},\ and\ \citenamefont
  {Parmigiani}}]{Crepaldi2012}%
  \BibitemOpen
  \bibfield  {author} {\bibinfo {author} {\bibfnamefont {A.}~\bibnamefont
  {Crepaldi}}, \bibinfo {author} {\bibfnamefont {B.}~\bibnamefont {Ressel}},
  \bibinfo {author} {\bibfnamefont {F.}~\bibnamefont {Cilento}}, \bibinfo
  {author} {\bibfnamefont {M.}~\bibnamefont {Zacchigna}}, \bibinfo {author}
  {\bibfnamefont {C.}~\bibnamefont {Grazioli}}, \bibinfo {author}
  {\bibfnamefont {H.}~\bibnamefont {Berger}}, \bibinfo {author} {\bibfnamefont
  {P.}~\bibnamefont {Bugnon}}, \bibinfo {author} {\bibfnamefont
  {K.}~\bibnamefont {Kern}}, \bibinfo {author} {\bibfnamefont {M.}~\bibnamefont
  {Grioni}},\ and\ \bibinfo {author} {\bibfnamefont {F.}~\bibnamefont
  {Parmigiani}},\ }\bibfield  {title} {\bibinfo {title} {Ultrafast photodoping
  and effective {Fermi-Dirac} distribution of the {Dirac} particles in
  {Bi$_2$Se$_3$}},\ }\href {https://doi.org/10.1103/physrevb.86.205133}
  {\bibfield  {journal} {\bibinfo  {journal} {Physical Review B}\ }\textbf
  {\bibinfo {volume} {86}},\ \bibinfo {pages} {205133} (\bibinfo {year}
  {2012})}\BibitemShut {NoStop}%
\bibitem [{\citenamefont {Crepaldi}\ \emph {et~al.}(2013)\citenamefont
  {Crepaldi}, \citenamefont {Cilento}, \citenamefont {Ressel}, \citenamefont
  {Cacho}, \citenamefont {Johannsen}, \citenamefont {Zacchigna}, \citenamefont
  {Berger}, \citenamefont {Bugnon}, \citenamefont {Grazioli}, \citenamefont
  {Turcu}, \citenamefont {Springate}, \citenamefont {Kern}, \citenamefont
  {Grioni},\ and\ \citenamefont {Parmigiani}}]{Crepaldi2013}%
  \BibitemOpen
  \bibfield  {author} {\bibinfo {author} {\bibfnamefont {A.}~\bibnamefont
  {Crepaldi}}, \bibinfo {author} {\bibfnamefont {F.}~\bibnamefont {Cilento}},
  \bibinfo {author} {\bibfnamefont {B.}~\bibnamefont {Ressel}}, \bibinfo
  {author} {\bibfnamefont {C.}~\bibnamefont {Cacho}}, \bibinfo {author}
  {\bibfnamefont {J.~C.}\ \bibnamefont {Johannsen}}, \bibinfo {author}
  {\bibfnamefont {M.}~\bibnamefont {Zacchigna}}, \bibinfo {author}
  {\bibfnamefont {H.}~\bibnamefont {Berger}}, \bibinfo {author} {\bibfnamefont
  {P.}~\bibnamefont {Bugnon}}, \bibinfo {author} {\bibfnamefont
  {C.}~\bibnamefont {Grazioli}}, \bibinfo {author} {\bibfnamefont {I.~C.~E.}\
  \bibnamefont {Turcu}}, \bibinfo {author} {\bibfnamefont {E.}~\bibnamefont
  {Springate}}, \bibinfo {author} {\bibfnamefont {K.}~\bibnamefont {Kern}},
  \bibinfo {author} {\bibfnamefont {M.}~\bibnamefont {Grioni}},\ and\ \bibinfo
  {author} {\bibfnamefont {F.}~\bibnamefont {Parmigiani}},\ }\bibfield  {title}
  {\bibinfo {title} {Evidence of reduced surface electron-phonon scattering in
  the conduction band of {Bi$_2$Se$_3$} by nonequilibrium {ARPES}},\ }\href
  {https://doi.org/10.1103/physrevb.88.121404} {\bibfield  {journal} {\bibinfo
  {journal} {Physical Review B}\ }\textbf {\bibinfo {volume} {88}},\ \bibinfo
  {pages} {121404} (\bibinfo {year} {2013})}\BibitemShut {NoStop}%
\bibitem [{sup()}]{supp}%
  \BibitemOpen
  \href@noop {} {}\bibinfo {note} {See Supplemental Material at (insert URL)
  for the calculations at intermediate SOC values and the procedure followed to
  determine the binding energy of the Dirac point.}\BibitemShut {Stop}%
\bibitem [{\citenamefont {Hsieh}\ \emph {et~al.}(2009)\citenamefont {Hsieh},
  \citenamefont {Xia}, \citenamefont {Qian}, \citenamefont {Wray},
  \citenamefont {Dil}, \citenamefont {Meier}, \citenamefont {Osterwalder},
  \citenamefont {Patthey}, \citenamefont {Checkelsky}, \citenamefont {Ong},
  \citenamefont {Fedorov}, \citenamefont {Lin}, \citenamefont {Bansil},
  \citenamefont {Grauer}, \citenamefont {Hor}, \citenamefont {Cava},\ and\
  \citenamefont {Hasan}}]{Hsieh2009a}%
  \BibitemOpen
  \bibfield  {author} {\bibinfo {author} {\bibfnamefont {D.}~\bibnamefont
  {Hsieh}}, \bibinfo {author} {\bibfnamefont {Y.}~\bibnamefont {Xia}}, \bibinfo
  {author} {\bibfnamefont {D.}~\bibnamefont {Qian}}, \bibinfo {author}
  {\bibfnamefont {L.}~\bibnamefont {Wray}}, \bibinfo {author} {\bibfnamefont
  {J.~H.}\ \bibnamefont {Dil}}, \bibinfo {author} {\bibfnamefont
  {F.}~\bibnamefont {Meier}}, \bibinfo {author} {\bibfnamefont
  {J.}~\bibnamefont {Osterwalder}}, \bibinfo {author} {\bibfnamefont
  {L.}~\bibnamefont {Patthey}}, \bibinfo {author} {\bibfnamefont {J.~G.}\
  \bibnamefont {Checkelsky}}, \bibinfo {author} {\bibfnamefont {N.~P.}\
  \bibnamefont {Ong}}, \bibinfo {author} {\bibfnamefont {A.~V.}\ \bibnamefont
  {Fedorov}}, \bibinfo {author} {\bibfnamefont {H.}~\bibnamefont {Lin}},
  \bibinfo {author} {\bibfnamefont {A.}~\bibnamefont {Bansil}}, \bibinfo
  {author} {\bibfnamefont {D.}~\bibnamefont {Grauer}}, \bibinfo {author}
  {\bibfnamefont {Y.~S.}\ \bibnamefont {Hor}}, \bibinfo {author} {\bibfnamefont
  {R.~J.}\ \bibnamefont {Cava}},\ and\ \bibinfo {author} {\bibfnamefont
  {M.~Z.}\ \bibnamefont {Hasan}},\ }\bibfield  {title} {\bibinfo {title} {A
  tunable topological insulator in the spin helical {Dirac} transport regime},\
  }\href {https://doi.org/10.1038/nature08234} {\bibfield  {journal} {\bibinfo
  {journal} {Nature}\ }\textbf {\bibinfo {volume} {460}},\ \bibinfo {pages}
  {1101} (\bibinfo {year} {2009})}\BibitemShut {NoStop}%
\bibitem [{\citenamefont {Aguilera}\ \emph {et~al.}(2019)\citenamefont
  {Aguilera}, \citenamefont {Friedrich},\ and\ \citenamefont
  {Blügel}}]{Aguilera2019}%
  \BibitemOpen
  \bibfield  {author} {\bibinfo {author} {\bibfnamefont {I.}~\bibnamefont
  {Aguilera}}, \bibinfo {author} {\bibfnamefont {C.}~\bibnamefont
  {Friedrich}},\ and\ \bibinfo {author} {\bibfnamefont {S.}~\bibnamefont
  {Blügel}},\ }\bibfield  {title} {\bibinfo {title} {Many-body corrected
  tight-binding {Hamiltonians} for an accurate quasiparticle description of
  topological insulators of the {Bi$_2$Se$_3$} family},\ }\href
  {https://doi.org/10.1103/physrevb.100.155147} {\bibfield  {journal} {\bibinfo
   {journal} {Physical Review B}\ }\textbf {\bibinfo {volume} {100}},\ \bibinfo
  {pages} {155147} (\bibinfo {year} {2019})}\BibitemShut {NoStop}%
\bibitem [{\citenamefont {Ponzoni}\ \emph {et~al.}(2023)\citenamefont
  {Ponzoni}, \citenamefont {Pa{\ss}lack}, \citenamefont {Stupar}, \citenamefont
  {Janas}, \citenamefont {Zamborlini},\ and\ \citenamefont
  {Cinchetti}}]{Ponzoni2023}%
  \BibitemOpen
  \bibfield  {author} {\bibinfo {author} {\bibfnamefont {S.}~\bibnamefont
  {Ponzoni}}, \bibinfo {author} {\bibfnamefont {F.}~\bibnamefont
  {Pa{\ss}lack}}, \bibinfo {author} {\bibfnamefont {M.}~\bibnamefont {Stupar}},
  \bibinfo {author} {\bibfnamefont {D.~M.}\ \bibnamefont {Janas}}, \bibinfo
  {author} {\bibfnamefont {G.}~\bibnamefont {Zamborlini}},\ and\ \bibinfo
  {author} {\bibfnamefont {M.}~\bibnamefont {Cinchetti}},\ }\bibfield  {title}
  {\bibinfo {title} {Dirac bands in the topological insulator {Bi$_2$Se$_3$}
  mapped by time-resolved momentum microscopy},\ }\href
  {https://doi.org/10.1002/apxr.202200016} {\bibfield  {journal} {\bibinfo
  {journal} {Advanced Physics Research}\ ,\ \bibinfo {pages} {2200016}}
  (\bibinfo {year} {2023})}\BibitemShut {NoStop}%
\bibitem [{\citenamefont {Boschini}\ \emph {et~al.}(2020)\citenamefont
  {Boschini}, \citenamefont {Bugini}, \citenamefont {Zonno}, \citenamefont
  {Michiardi}, \citenamefont {Day}, \citenamefont {Razzoli}, \citenamefont
  {Zwartsenberg}, \citenamefont {Schneider}, \citenamefont {da~Silva~Neto},
  \citenamefont {dal Conte}, \citenamefont {Kushwaha}, \citenamefont {Cava},
  \citenamefont {Zhdanovich}, \citenamefont {Mills}, \citenamefont {Levy},
  \citenamefont {Carpene}, \citenamefont {Dallera}, \citenamefont {Giannetti},
  \citenamefont {Jones}, \citenamefont {Cerullo},\ and\ \citenamefont
  {Damascelli}}]{Boschini2020}%
  \BibitemOpen
  \bibfield  {author} {\bibinfo {author} {\bibfnamefont {F.}~\bibnamefont
  {Boschini}}, \bibinfo {author} {\bibfnamefont {D.}~\bibnamefont {Bugini}},
  \bibinfo {author} {\bibfnamefont {M.}~\bibnamefont {Zonno}}, \bibinfo
  {author} {\bibfnamefont {M.}~\bibnamefont {Michiardi}}, \bibinfo {author}
  {\bibfnamefont {R.~P.}\ \bibnamefont {Day}}, \bibinfo {author} {\bibfnamefont
  {E.}~\bibnamefont {Razzoli}}, \bibinfo {author} {\bibfnamefont
  {B.}~\bibnamefont {Zwartsenberg}}, \bibinfo {author} {\bibfnamefont
  {M.}~\bibnamefont {Schneider}}, \bibinfo {author} {\bibfnamefont {E.~H.}\
  \bibnamefont {da~Silva~Neto}}, \bibinfo {author} {\bibfnamefont
  {S.}~\bibnamefont {dal Conte}}, \bibinfo {author} {\bibfnamefont {S.~K.}\
  \bibnamefont {Kushwaha}}, \bibinfo {author} {\bibfnamefont {R.~J.}\
  \bibnamefont {Cava}}, \bibinfo {author} {\bibfnamefont {S.}~\bibnamefont
  {Zhdanovich}}, \bibinfo {author} {\bibfnamefont {A.~K.}\ \bibnamefont
  {Mills}}, \bibinfo {author} {\bibfnamefont {G.}~\bibnamefont {Levy}},
  \bibinfo {author} {\bibfnamefont {E.}~\bibnamefont {Carpene}}, \bibinfo
  {author} {\bibfnamefont {C.}~\bibnamefont {Dallera}}, \bibinfo {author}
  {\bibfnamefont {C.}~\bibnamefont {Giannetti}}, \bibinfo {author}
  {\bibfnamefont {D.~J.}\ \bibnamefont {Jones}}, \bibinfo {author}
  {\bibfnamefont {G.}~\bibnamefont {Cerullo}},\ and\ \bibinfo {author}
  {\bibfnamefont {A.}~\bibnamefont {Damascelli}},\ }\bibfield  {title}
  {\bibinfo {title} {Role of matrix elements in the time-resolved photoemission
  signal},\ }\href {https://doi.org/10.1088/1367-2630/ab6eb1} {\bibfield
  {journal} {\bibinfo  {journal} {New Journal of Physics}\ }\textbf {\bibinfo
  {volume} {22}},\ \bibinfo {pages} {023031} (\bibinfo {year}
  {2020})}\BibitemShut {NoStop}%
\end{thebibliography}
\end{document}



\title{Supplementary Material for: Unveiling the physics of the spin-orbit coupling at the surface of a model topological insulator: from theory to experiments}

\author{D.~Puntel} 
 \affiliation{Dipartimento di Fisica, Università degli Studi di Trieste, 34127 Trieste, Italy}

 \author{S.~Peli}
 \affiliation{Elettra - Sincrotrone Trieste S.C.p.A., Strada Statale 14, km 163.5, 34149 Trieste, Italy}
 
 \author{W.~Bronsch}
 \affiliation{Elettra - Sincrotrone Trieste S.C.p.A., Strada Statale 14, km 163.5, 34149 Trieste, Italy}
 
\author{F.~Cilento}%
\affiliation{Elettra - Sincrotrone Trieste S.C.p.A., Strada Statale 14, km 163.5, 34149 Trieste, Italy}%

 \author{H.~Ebert}
 \affiliation{Department Chemie, Ludwig-Maximilians-University M\"unchen, Butenandtstr. 5-11, 81377 München, Germany}
 
 \author{J.~Braun}
 \affiliation{Department Chemie, Ludwig-Maximilians-University M\"unchen, Butenandtstr. 5-11, 81377 München, Germany}
 
 \author{F.~Parmigiani}
 \email{fulvio.parmigiani@elettra.eu}
 \affiliation{Dipartimento di Fisica, Università degli Studi di Trieste, 34127 Trieste, Italy}
 \affiliation{Elettra - Sincrotrone Trieste S.C.p.A., Strada Statale 14, km 163.5, Trieste, Italy}
 \affiliation{International Faculty, University of Cologne, Albertus-Magnus-Platz, 50923 Cologne, Germany}
 
 \maketitle

\section{Photoemission calculations for 100\%, 85\%, 70\%, 50\% surface SOC}

\begin{figure}[h!]
\centering
\includegraphics[width=0.9\columnwidth]{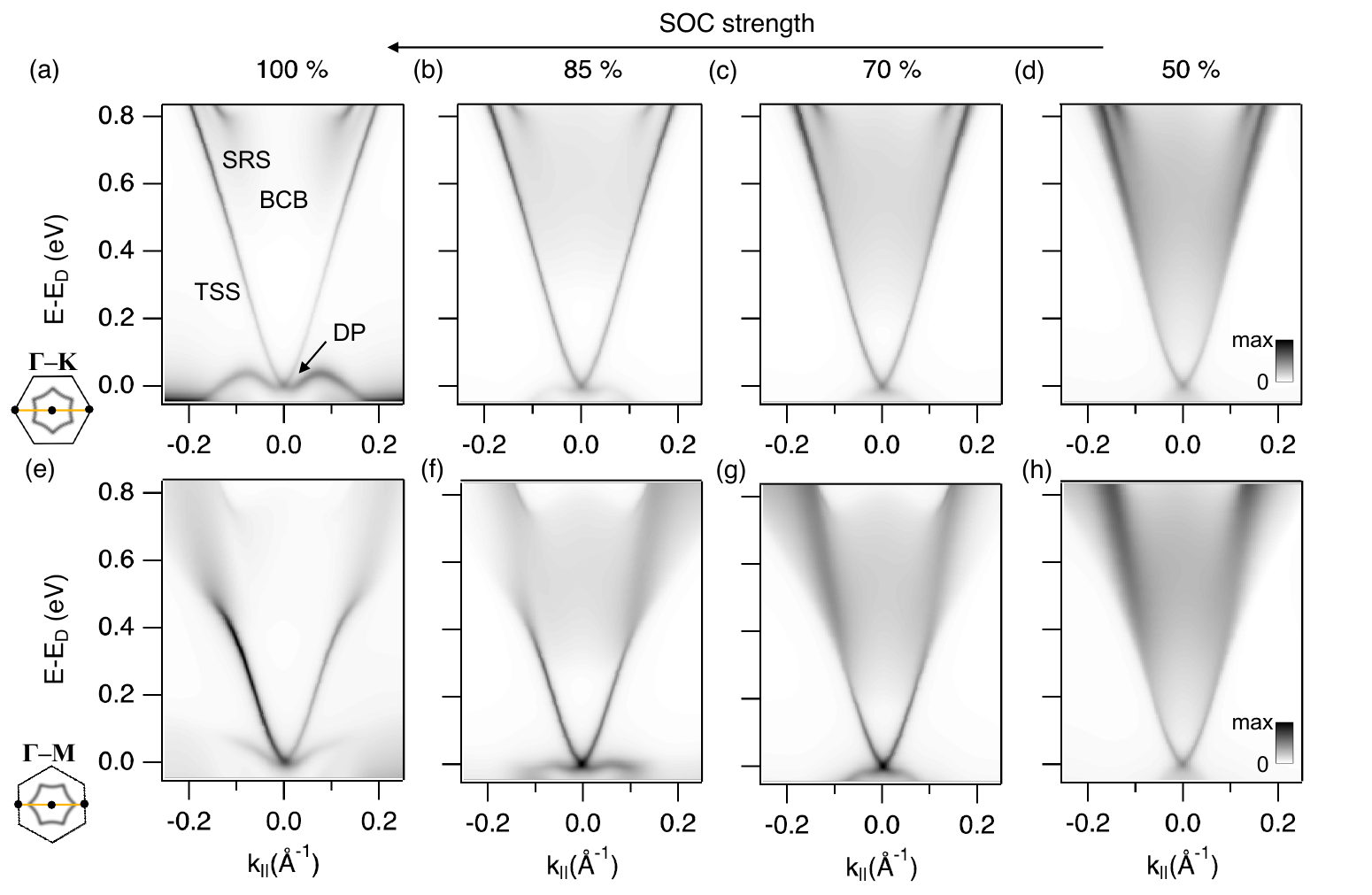}
\caption{(a-d) Calculated \emph{E~vs.~k} photoemission intensity along the $\Gamma$K direction (as indicated in the lower left part of panel (a)) for $s$-polarized photons of 10.8\,eV, upon varying the SOC strength from 100\% (a) to 50\% (d). Panel (a) also indicates the features relevant to our study: TSS, topological surface state; BCB, bulk conduction band; SRS, surface resonance state; DP, Dirac point, to which the energy scale is referenced. (e-h) Same as (a-d), along the $\Gamma$M direction. 
}
\label{Fig_S1}
\end{figure}

Additional calculations for the bandstructure of Bi$_2$Se$_3$ performed at intermediate values of the surface SOC, as referenced to in the main text, are shown in Fig. \ref{Fig_S1}. We report the cases of 100\%, 85\%, 70\% and 50\% SOC for the $\Gamma$K (a-d) and the $\Gamma$M (e-h) direction. Panels (a), (d), (e) and (h) constitute Fig.~2 of the main text. The relevant features for the discussion in the main text are indicated in panel (a). Note, for both directions, the gradual smearing-out of the ``M''-like shape around the Dirac point as the SOC is reduced from 100\%, persisting at 85\% SOC but no longer at 70\% SOC. Due to the contrast with the topological surface state (TSS) and the bulk conduction band (BCB), along $\Gamma$M the surface resonance state (SRS) is more visible at 85\% SOC than at the other values. 

\pagebreak

\section{Determination of the Dirac point binding energy}

\begin{figure}[h!]
\centering
\includegraphics[width=0.9\columnwidth]{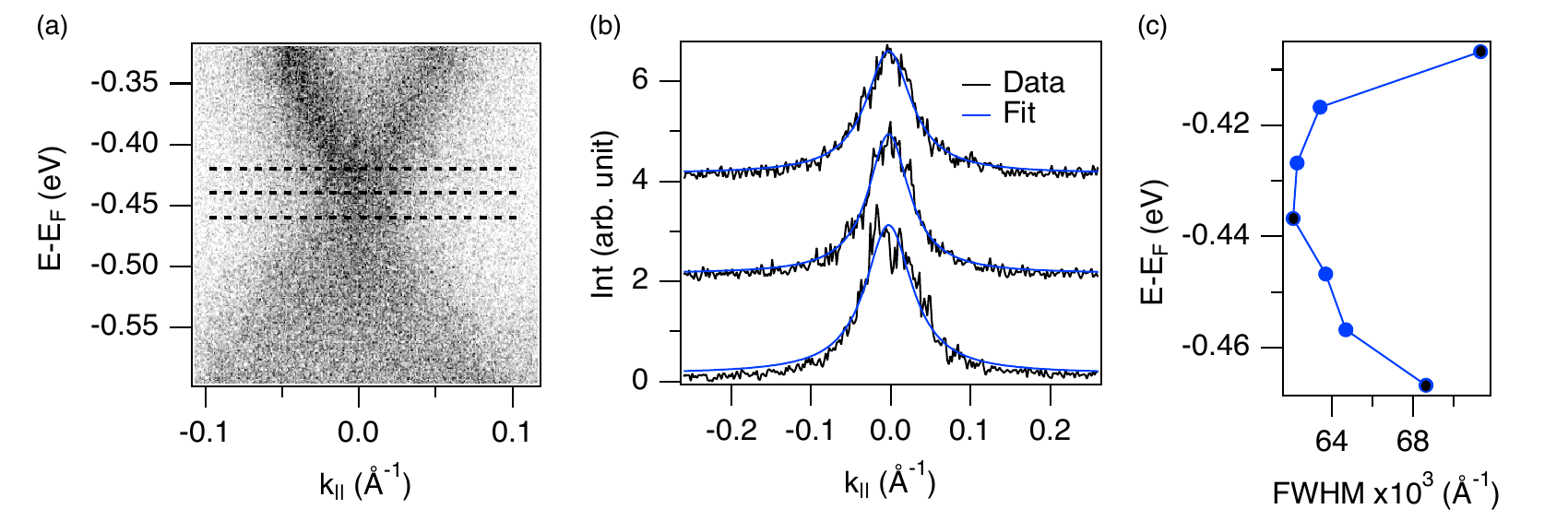}
\caption{(a) Dispersion close to the Dirac point, measured along $\Gamma$M with $s$-polarized photons. (b) Selected MDC cuts taken in correspondence of the black dashed lines in panel (a). The blue line is the fit of the data to a Gaussian-broadened Lorentzian peak. (c) Lorentzian FHWM of the MDC cuts extracted from the fit, as a function of the binding energy. The values for the cuts reported in panel (b) are overmarked in black.
}
\label{Fig_S2}
\end{figure}

We illustrate the procedure followed to determine the binding energy of the Dirac point, as outlined in the main text. Figure~\ref{Fig_S2}(a) shows an equilibrium measurement along $\Gamma$M with $s$-polarized photons, in the vicinity of the Dirac point. In order to identify the binding energy of the Dirac point, we extracted momentum distribution curves (MDC) integrating the intensity over 10\,meV-windows above and below the estimated Dirac point energy. The black dashed lines in panel (a) show the central position of three of them. Owing to the convergence of the two TSS branches, we fit them with a single Lorentzian peak, with Gaussian broadening to account for the experimental resolution. The result is reported in Fig.~\ref{Fig_S2}(b) for the three cuts indicated in (a), along with the corresponding fit in blue. The Dirac point is assumed to be located at the energy where the MDC of minimum width has been extracted. The width obtained from the fit is shown in Fig.\ref{Fig_S2} as FWHM of the Lorentzian peak \emph{versus} the binding energy, for all the seven MDCs we extracted. The three representative cuts already reported in panel (a) and (b) are here marked by an additional black marker. The lowest-width one, at $\sim$-0.44\,eV, is identified as the Dirac point. 

